\documentclass[fleqn,usenatbib]{mnras}

\usepackage{newtxtext,newtxmath}

\usepackage[T1]{fontenc}

\DeclareRobustCommand{\VAN}[3]{#2}
\let\VANthebibliography\thebibliography
\def\thebibliography{\DeclareRobustCommand{\VAN}[3]{##3}\VANthebibliography}

\usepackage{graphicx}	
\usepackage{amsmath}	

\title[$Q$ for discs of stars and gas]{Redefining \boldmath $Q$ for multi-component discs of stars and gas}

\author[K. George et al.]{
Kit George,$^{1}$\thanks{E-mail:kit.george.21@ucl.ac.uk}
Ralph Schönrich$^{1}$
\\
$^{1}$Mullard Space Science Laboratory, University College London, Holmbury St Mary, Dorking, Surrey, RH5 6NT, UK\\
}

\date{Accepted XXX. Received YYY; in original form ZZZ}

\pubyear{2025}

\begin{document}
\label{firstpage}
\pagerange{\pageref{firstpage}--\pageref{lastpage}}
\maketitle

\begin{abstract}
We point out a fundamental mismatch in the $Q$ stability parameter for Galactic discs: Toomre's $Q = 1$ defines the boundary between axisymmetric stability/instability, while simulations, observations, and theoretical expectations apply $Q$ in the region $Q > 1$ as a measure for spiral activity (e.g. swing amplification), for which $Q$ has not been designed. We suggest to redefine $Q$ to keep $Q = 1$ as the stability boundary, but to equally yield a consistent map between $Q$ and the maximum swing amplification factor. Using the Goldreich-Lynden-Bell formalism, we find that particularly the $Q$ for gas discs has been mismatched, and should be redefined to close to the square of the traditional definition. We provide new formulations of $Q$ for simple, two-component, and multi-component discs, including a discussion of vertically extended discs, providing a simple iterative formula for which we also provide code. We find $Q \approx 1.58$ for the Solar Neighbourhood under our definition, closer to results from simulations. We compare the Milky Way and M74, showing that,  consistent with observations, the theory suggests a higher $m$ number for the Milky Way (arguing against a 2-arm pattern) for stellar-dominated patterns. Gas instability arises at much smaller scales ($m \gtrsim 10$), and we link both M74's gas pattern and local spurs in the Milky Way to this gas instability rather than stellar spiral arms.
\end{abstract}



\section{Introduction}
At the centre of galaxy research is the question of how discs' structure is shaped by secular instability: substructures like bars or spiral patterns naturally arise when discs become sufficiently massive and cold (i.e. low velocity dispersion). Discs typically form new stars on near circular orbits, which drives new instability \citep{SellwoodCarlberg1984}, so they will in most cases be at the brink of instability, where the heating from substructure is balanced by fresh star formation that drives new instability. Consistent with this general expectation, it was noted in the middle of the last century e.g. by \citet{Lin1961}, and \citet{Toomre1964} that in particular the Milky Way's Galactic disc appears to be at the margin of stability, as evidenced by its rich substructure like its spiral patterns and the  central Galactic bar (established a bit later by \citealt{Shane1971}) and its spiral patterns, whose precise topology is still under debate. 
Similar to the topology, our knowledge about mechanisms behind spiral arm formation and propagation is still imperfect. Bars (e.g. \citealt{PeschkenLokas2018}) and particularly grand-design spiral patterns \citep{Holmberg1941, PfleidererSiedentopf1961, Kazantzidis2009} can be created or driven by companion/satellite galaxies, see e.g. the case of M51 \citep{ToomreToomre1972, SaloLaurikainen2000, Dobbsea2010}. Even in the absence of external perturbations, numerical studies have shown that a kinematically cool disc will quickly develop spiral patterns \citep{Hohl1971, SellwoodCarlberg1984}, which will heat the disc (by scattering stars) until it lacks sufficient gravitational instability to sustain them \citep{SellwoodCarlberg1984}. Gravitational stability typically corresponds, locally, to the $Q$ parameter, or the radial velocity dispersion of stars in the disc divided by its surface density \citep{Toomre1964}.

The precise nature of spiral instabilities is still under debate. One mechanism is the groove instability \citep{SellwoodLin1989, SellwoodKahn1991}. An axisymmetric groove in the distribution function (DF) of stars, that is a narrow underdensity at a certain radius, in a kinematically cool disc can give rise to a continuously growing spiral. Eventually the spiral becomes strong enough to perturb the orbits of stars at its inner and outer Lindblad resonances, creating groove-like features at these radii on which new spirals can grow \citep{SellwoodCarlberg2019}. As a result, at any given time there may be multiple local spiral patterns active at different radii in the disc, creating the appearance of a global shearing pattern as discussed in \citet{SellwoodCarlberg2014}. The groove instability is understood as a linear instability, but it has been argued by e.g. \citet{Donghia2013} that since spiral patterns become non-linear, linear theory cannot provide meaningful understanding after the initial growth phase.

A mechanism that appears to underlie linear gravitational instabilities (e.g. \citealt{SellwoodKahn1991}) and may also contribute to the non-linear behaviour of spirals is swing amplification: the notion that shearing spiral patterns increase in strength as they swing from leading to trailing. \citet{Toomre1981}, using the formalism of \citet{GoldreichLynden-Bell1965} but modified with a stellar reduction factor, calculated the local linear evolution of a leading spiral wave as it sheared to trailing alignment, and demonstrated that the amplification of the pattern's strength was due to the interaction between the retrograde rotation of the spiral pattern's shear and the retrograde epicyclic oscillations of individual stars. This calculation does not generalise to arbitrary patterns in position-velocity phase space, so it cannot be used to describe the evolution of real spiral patterns in terms of swing amplification, but it approximately agrees with equivalent calculations in the more complete (also local and linear) formalism of \citet{JulianToomre1966} \citep{BinneyTremaine2008}. \citet{MeidtVanderWel2024} provides a more general formalism for calculating the evolution of patterns under swing amplification, which may be tested against spiral patterns observationally and in simulations. Whether the gravitational instabilities found in simulations can legitimately be called swing amplification remains an interesting qualitative question \citep{SellwoodKahn1991, Donghia2013} but the calculation in \citet{GoldreichLynden-Bell1965} and \citet{Toomre1981} remains a useful indicative measure of a disc's instability with respect to spiral patterns. 

We still do not know reliably the number of Galactic spiral arms, though the majority of investigations point to the Milky Way having a four-armed spiral pattern (e.g. \citealt{BobylevBajkova2014, ShenZheng2020}). An interesting complication of spiral patterns is the existence of minor arm features, called "spurs" or "feathers", which branch off the main spiral arms \citep{Elmegreen1980, LaVigneea2006}. It has been suggested that these are a result of gravitational instability in the stars \citep{Elmegreen1980, Byrdea1984} or in the interstellar gas \citep{Balbus1988, KimOstriker2002}. The former describe stellar instability under the formalism in \citet{JulianToomre1966} while the latter use an extension of the formalism in \citet{GoldreichLynden-Bell1965} to describe instability in the gas around a growing spur. The comparison of these explanations for the spurs, presented in \citet{Dobbsea2010}, is inconclusive. We will demonstrate later how the typical length-scale of such an instability differs between the stellar and gas components of a disc.

Throughout this paper we will apply the formalism of \citet{GoldreichLynden-Bell1965} and \citet{Toomre1981} (referred to as the GLB equation) to discs composed of stars and gas, as a measure of their relative instabilities with respect to spiral structure. While existing analyses are more rigorous in the exclusive treatment of either stars or gas \citep{Balbus1988, MeidtVanderWel2024}, this approach is indicative of the relative contributions of each component to spiral instabilities. In particular, this approach will allow us to evaluate and compare single-parameter descriptions of the stability of multi-component discs (extensions of Toomre's $Q$).
\section{Toomre's \boldmath ${Q}$: Existing definitions} 
As outlined above, $Q$ somewhat problematically quantifies two different kinds of instability: it was devised for the stability of a disc to axisymmetric density perturbations, and is more commonly used to parametrise the susceptibility of a disc to the growth of (non-axisymmetric) spiral perturbations, e.g. by swing amplification. For a simple stellar disc (e.g. with a fixed isothermal/Schwarzschild velocity distribution) the problem is not apparent, as two different (but fixed) values on the scale of $Q$ delineate axisymmetric instability and swing amplification respectively. The problem will become apparent when the shape of the distribution function varies and shifts the stability of swing/spiral amplification on this scale, which means that below we will have to discuss $Q$ for mixed populations. Similarly, there are problems for mixtures of gas and stars. 
\subsection{Toomre's \boldmath ${Q}$ for a disc of stars or gas}
\citet{Safronov1960} showed that a disc of gas (a disc treated as a collisional fluid) can be unstable under self-gravity to axisymmetric density perturbations. Locally the disc is stable to axisymmetric perturbations exactly if the following criterion is fulfilled \citep{BinneyTremaine2008}:
\begin{align}
    Q_\mathrm{g} = \frac{\kappa c}{\pi G \Sigma} \geq 1,
\end{align}
where $\kappa$ is the local in-plane epicycle frequency, $\Sigma$ is the local gas surface density and $c$ is the local sound speed of the gas. "Global" stability with respect to axisymmetric perturbations is set when the disc is locally stable at all radii. \citet{Toomre1964} showed, using the collisionless Boltzmann equation, that the analogous local criterion for a stellar disc with a Schwarzschild distribution function of velocity dispersion $\sigma_R$ is
\begin{align}
    Q_\mathrm{s} = \frac{\kappa \sigma_R}{3.36 G \Sigma} \geq 1.
\end{align}

\citet{Toomre1981} described the growth of spiral waves in a galactic disc by swing amplification with a local calculation, approximating the motion of the stars or gas fluid elements as oscillators whose spring rates are determined by self-gravity with their neighbours. He recovered the behaviour of a gas disc from \citet{GoldreichLynden-Bell1965}, and extended it to stellar discs. They showed that, for a given radial/azimuthal frequency ratio $\kappa/\Omega$ (which sets the rate of shearing between orbits at different radii), the maximum amplification a leading spiral wave (of some optimal azimuthal wavenumber $k_y$) can experience is a decreasing function of $Q$. They did not, however, compare the maximum amplification as a function of $Q_\mathrm{s}$ for stars and of $Q_\mathrm{g}$ for gas.

Both stellar and fluid/gas descriptions of instability are derived from the local dispersion relations under self-gravity in the WKB (tight-winding) approximation \citep{BinneyTremaine2008}, and turn out to be one-dimensional problems. For fluid discs we have
\begin{align}
\begin{split}
    s^2 &= 1 - \frac{2 \pi G \Sigma k}{\kappa^2} + \frac{c^2 k^2}{\kappa^2} \\
    &= 1 - 2 \pi G \Sigma \frac{k}{\kappa^2} + (\kappa c)^2 \left(\frac{k}{\kappa^2}\right)^2,
\end{split}
\end{align}
where $k$ is the magnitude of they total wavenumber (including radial and azimuthal components), and for stellar discs
\begin{align}
    s^2 &= 1 - 2 \pi G \Sigma \frac{k}{\kappa^2} \mathcal{F} (s, \chi) \label{eq:1c_stellar_dr}, \\
    \chi &= \frac{k^2 \sigma_R^2}{\kappa^2} = (\kappa \sigma_R)^2 \left(\frac{k}{\kappa^2}\right)^2.
\end{align}
Here, 
\begin{align}
    s = \frac{\omega - m \Omega}{\kappa}
\end{align}
is the relative beat frequency of an $m$-armed spiral wave with rest-frame frequency $\omega$ in the corotating frame against a star/fluid element moving on an orbit of azimuthal and radial frequencies $\Omega$ and $\kappa$. If e.g. $\omega =  m \Omega$, the star keeps the same phase against the pattern; if $s = 1$ the star will encounter one wavecrest per epicycle. The "stellar reduction factor" $\mathcal{F}$, "the factor by which the response of
the disk to a given spiral perturbation is reduced below the value for a cold
disk" \citep{BinneyTremaine2008}, is 
\begin{align}
    \mathcal{F} (s, \chi) &= \frac{2}{\chi} (1 - s^2) \mathrm{e}^{-\chi}\sum_{n = 1}^\infty \frac{I_n (\chi)}{1 - s^2/n^2} \label{eq:ReductionFactor},
\end{align}
where $I_n$ is the modified Bessel function of the first kind. The above can be found in \citet{BinneyTremaine2008}. 

We will now show that both axisymmetric and spiral instabilities depend only on a single disc parameter, $Q$. We note that the dispersion relations are invariant under the following two maps (which also leave $Q$ unchanged): 
\begin{align}
\begin{split}
    \Sigma &\rightarrow A \Sigma \\
    \kappa c &\rightarrow A \kappa c \; \mathrm{or} \; \kappa \sigma_R \rightarrow A \kappa \sigma_R \\
    k &\rightarrow k/A
\end{split}
\end{align}
for $A \in \mathbb{R}$, and
\begin{align}
\begin{split}
    \kappa &\rightarrow B \kappa \\
    c &\rightarrow c/B \; \mathrm{or} \; \sigma_R \rightarrow \sigma_R/B \\
    k &\rightarrow B^2 k.
\end{split}
\end{align}
for $B \in \mathbb{R}$. As an aside, the first mapping implies that a disc with the same $Q$ value but lower $\Sigma$ and $\sigma$ will display instability at a larger wavenumber $k$ (or respectively $m$). This is the reason behind maximal discs showing smaller-$m$ spiral activity than halo-dominated low-mass discs, as found empirically in \citet{CarlbergFreedman1985}.
The disc is locally unstable to axisymmetric ($m = 0$) perturbations if $s^2 < 0$ for some $k$, so that axisymmetric modes of that $k$ have an imaginary frequency $\omega$ so can grow exponentially. The swing amplification calculation from \citet{Toomre1981} is also based on these dispersion relations. $\min_k s^2$ (which determines axisymmetric stability) and Toomre's $\max_{k_y}(\mathrm{Amplification})$ are both invariant under the above maps for fixed shearing $\kappa/\Omega$, so out of our three degrees of freedom ($\kappa, \Sigma, c \, \mathrm{or} \, \sigma_R$) only one remains relevant to stability, $\kappa c / \Sigma$ or $\kappa \sigma_R / \Sigma$, which in either case is $Q$ up to a prefactor. Therefore, both axisymmetric stability and swing amplification can be described under a single parameter. Throughout this paper we will treat $\kappa$ as a constant. 
\subsection{Extending \boldmath ${Q}$ to multi-component discs} \label{sec:intro_extending_Q}
Several attempts have been made in the literature to define a $Q$ value for a more realistic disc of multiple components containing stars and gas \citep{WangSilk1994, Jog1996, BertinRomeo1988, RomeoWiegert2011, Romeo2013}. However, the stability of a multi-component disc is not a one-dimensional problem, so we must define carefully what we require of an extended definition for $Q$. \citet{Rafikov2001} found the WKB dispersion relation for a disc containing $n_\mathrm{g}$ components of gas and $n_\mathrm{s}$ components of stars (each stellar component having a Schwarzschild DF):
\begin{align}
    2 \pi G k& \left(\sum_{i = 1}^{n_\mathrm{g}} \frac{\Sigma_{\mathrm{g}, i}}{\kappa^2(1 - s^2) + k^2 c_i^2} + \sum_{j = 1}^{n_\mathrm{s}} \frac{\Sigma_{\mathrm{s}, j} \mathcal{F} (s, \chi_j)}{\kappa^2(1 - s^2)}\right) = 1, \label{eq:DispersionRelationRk} \\
    \chi_j &= \frac{k^2 \sigma_{R, j}^2}{\kappa^2}.
\end{align}
$s^2$ is again invariant under the map (now keeping $\kappa$ constant):
\begin{align}
\begin{split}
    \{\Sigma_{\mathrm{g}, i}\} &\rightarrow \{A \Sigma_{\mathrm{g}, i}\} \\
    \{\Sigma_{\mathrm{s}, j}\} &\rightarrow \{A \Sigma_{\mathrm{s}, j}\} \\
    \{c_i\} &\rightarrow \{A c_i\} \\
    \{\sigma_{R, j}\} &\rightarrow \{A \sigma_{R, j}\} \\
    k &\rightarrow k/A.
\end{split}
\end{align}
This leaves axisymmetric stability and maximum swing amplification each as functions of $2(n_\mathrm{g} + n_\mathrm{s}) - 1$ parameters, describing the relative values of the surface densities and sound speeds/velocity dispersions of the components, without specifying their magnitudes. These parameters might be the single-component $Q$ values of the individual components and their relative surface densities, or their relative surface densities, their relative sound speeds/velocity dispersions and the $Q$ value of one component. Thus, to extend the definition of $Q$ to a multi-component disc, we seek a function $Q$ of these $2(n_\mathrm{g} + n_\mathrm{s}) - 1$ parameters which fulfills the following:
\begin{enumerate}
    \item $Q \geq 1 \iff$ disc is stable to axisymmetric perturbations
    \item The maximum swing amplification factor is a decreasing, one-to-one function of $Q$
    \item $Q$ reduces to the standard definition for a single-component stellar disc
\end{enumerate}

Before discussing existing definitions for $Q$ in multi-component discs we introduce common notation for describing the axisymmetric stability criterion. The disc is stable to axisymmetric perturbations if $s^2 > 0 \, \forall \, k$ according to the dispersion relation (\ref{eq:DispersionRelationRk}). Equivalently:
\begin{align}
\begin{split}
    F_\mathrm{max} &(\{\Sigma_{\mathrm{g}, i}\}, \{\Sigma_{\mathrm{s}, j}\}, \{c_i\}, \{\sigma_{R, j}\}) \\
    = &\max_k F (\{\Sigma_{\mathrm{g}, i}\}, \{\Sigma_{\mathrm{s}, j}\}, \{c_i\}, \{\sigma_{R, j}\}, k) \leq 1, \end{split} \\
\begin{split}    
F(\{\Sigma&_{\mathrm{g}, i}\}, \{\Sigma_{\mathrm{s}, j}\}, \{c_i\}, \{\sigma_{R, j}\}, k) \\
    = {}& 2 \pi G k \left(\sum_{i = 1}^{n_\mathrm{g}} \frac{\Sigma_{\mathrm{g},i}}{\kappa^2 + k^2 c_i^2} + \sum_{j = 1}^{n_\mathrm{s}} \frac{\Sigma_{\mathrm{s},j}}{\kappa^2} \mathcal{F}(0, \chi_j)\right) \label{eq:F_definition}.
    \end{split}
\end{align}
Consider a single-component fluid disc (though this argument also applies to a single component stellar disc). The axisymmetric stability condition
\begin{align}
    F_\mathrm{max}(\Sigma, c) \leq 1
\end{align}
can also be written
\begin{align}
    Q &= \frac{\kappa c}{\pi G \Sigma} \geq 1,
\end{align}
so if we map $c \rightarrow c/Q$ we obtain a disc which is marginally stable to axisymmetric perturbations; we can always write
\begin{align}
    \frac{\kappa}{\pi G \Sigma} \frac{c}{Q} = 1,
\end{align}
or equivalently
\begin{align}
    F_\mathrm{max}(\Sigma, \frac{c}{Q}) = 1 \label{eq:1cMarginal}.
\end{align}
Writing the definition for $Q$ in this notation can be convenient for extensions of the definition to multi-component discs of stars and gas, as we shall see in the next subsection and throughout this paper.
\subsection{Existing multi-component \boldmath ${Q}$ definitions}
\label{sec:existing_definitions}

\begin{figure*}
    \centering
    \includegraphics[width=\textwidth]{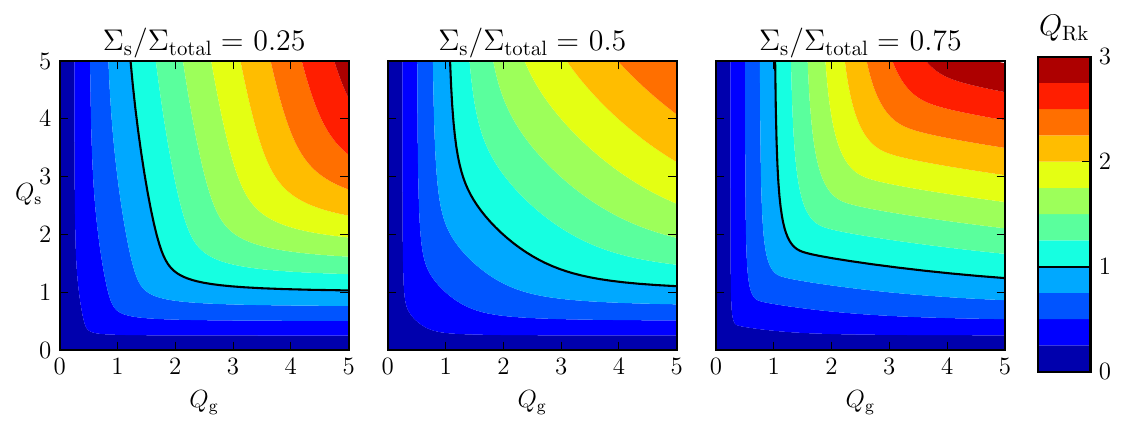}
    \caption{Contours of constant $Q_\mathrm{Rk}$ (equation \ref{eq:Q_Rk_definition}) for a 2-component disc of stars and gas with individual $Q$ values for its gas and stellar components $Q_\mathrm{g}$ and $Q_\mathrm{s}$ respectively, for various mass ratios between the two components. Each contour is an enlargement of the $Q_\mathrm{Rk} = 1$ contour about the origin by factor $Q_\mathrm{Rk}$.}
    \label{fig:q_rk_contours}
\end{figure*}

\citet{BertinRomeo1988} extended equation (\ref{eq:1cMarginal}) to define $Q$ for a disc of two fluid (i.e. gas) components using the two-fluid WKB dispersion relation from \citet{JogSolomon1984}. In our notation, their definition reads
\begin{align}
    F_\mathrm{max}(\Sigma_\mathrm{1}, \Sigma_\mathrm{2}, \frac{c_1}{Q_\mathrm{BR}}, \frac{c_2}{Q_\mathrm{BR}}) = 1.
\end{align}
As discussed for the dispersion relation in Section \ref{sec:intro_extending_Q}, $F_\mathrm{max}$ is invariant under multiplication of all its arguments by a constant:
\begin{align}
    F_\mathrm{max}(Q_\mathrm{BR} \Sigma_1, Q_\mathrm{BR} \Sigma_2, c_1, c_2) = 1 .\label{eq:Q_BR_multiplied_densities}
\end{align}
From equation (\ref{eq:F_definition}) we have
\begin{align}
    F(\Sigma_1, \Sigma_2, c_1, c_2, k) &= \frac{1}{Q_\mathrm{BR}} F(Q_\mathrm{BR} \Sigma_1, Q_\mathrm{BR} \Sigma_2, c_1, c_2, k),
\end{align}
and so
\begin{align}
    F_\mathrm{max}(\Sigma_1, \Sigma_2, c_1, c_2) = \frac{1}{Q_\mathrm{BR}} F_\mathrm{max}(Q_\mathrm{BR} \Sigma_1, Q_\mathrm{BR} \Sigma_2, c_1, c_2) = \frac{1}{Q_\mathrm{BR}}.
\end{align}
For a multi-component disc of stars and gas \citet{Romeo2013} and others (e.g. \citealt{LiMacLowKlessen2005}) defined $Q$ directly from the dispersion relation (\ref{eq:DispersionRelationRk}) of \citet{Rafikov2001}:
\begin{align}
    \frac{1}{Q_\mathrm{Rk}} = F_\mathrm{max}(\{\Sigma_{\mathrm{g}, i}\}, \{\Sigma_{\mathrm{s}, j}\}, \{c_i\}, \{\sigma_{R, j}\}). \label{eq:Q_Rk_definition}
\end{align}
For calculational convenience \citet{Romeo2013} and others (e.g. \citealt{LeroyEa2008}) defined a $Q$ value which instead treated all components as a fluid (which is simply an extension of $Q_\mathrm{BR}$ to $N$ components):
\begin{align}
    \frac{1}{Q_\mathrm{Rf}} = F_\mathrm{max}(\{\Sigma_i\}, \{c_i\}).
\end{align}
These three definitions ($Q_{\mathrm{BR}}$, $Q_{\mathrm{Rk}}$, $Q_{\mathrm{Rf}}$) are all direct extensions of equation (\ref{eq:1cMarginal}). Figure \ref{fig:q_rk_contours} plots $Q_\mathrm{Rk}$ for two-component discs of stars and gas with various surface density ratios against the individual $Q$ values of the components, $Q_\mathrm{g}$ and $Q_\mathrm{s}$. Note that each contour is an enlargement of the $Q_\mathrm{Rk} = 1$ contour about the origin by factor $Q_\mathrm{Rk}$.

\citet{RomeoWiegert2011} approximated $Q_\mathrm{BR}$ by treating a disc of stars and gas as a two-fluids system with the analytical expression:
\begin{align}
    Q_\mathrm{RW} &= 
    \begin{cases}
        \frac{W}{Q_1} + \frac{1}{Q_2} & Q_\mathrm{1} \geq Q_\mathrm{2} \\
        \frac{1}{Q_1} + \frac{W}{Q_2} & Q_\mathrm{2} \geq Q_\mathrm{1}
    \end{cases}  , {\mathrm{with }} \,\,\, W = \frac{2 c_1 c_2}{c_1^2 + c_2^2} .
\end{align}
Extending this, \citet{Romeo2013} showed that $Q_\mathrm{RW}$ can approximate an $N$-component disc with all components treated as fluids:
\begin{align}
    Q_\mathrm{RW} &= \sum_{i = 1}^{N} \frac{W_i}{Q_i}   , {\mathrm{with }} \,\,\, W_i = \frac{2 c_m c_i}{c_m^2 + c_i^2}.
\end{align}
Here $c_m$ is the sound speed of the component with the lowest individual $Q$:  $Q_m = \min \{Q_i\}$. However, \citet{Romeo2013} derived an upper bound for the error on this value compared to $Q_\mathrm{Rf}$ of
\begin{align}
   \frac{\langle \Delta Q_{\mathrm{RW}} \rangle} {Q_{\mathrm{Rf}}} \approx 0.03 \sqrt{N},
\end{align}
implying that $Q_\mathrm{RW}$ is unsuitable for discs of many ($N$) components. $Q_\mathrm{RW}$ is an adaptation of the $Q$ value from \citet{WangSilk1994} for a two-fluid disc:
\begin{align}
    \frac{1}{Q_\mathrm{WS}} = \frac{1}{Q_1} + \frac{1}{Q_2},
\end{align}
which leaned on the two-fluid dispersion relation in \citet{JogSolomon1984}, but was later shown to be erroneous by \citet{Jog1996}, who observed that for a single-component fluid disc
\begin{align}
    Q^2 = \frac{2}{F(\Sigma, c, k_\mathrm{min})},
\end{align}
where $k_\mathrm{min}$ is the value of $k$ which minimises $s^2$ in the dispersion relation, and extended this to define a $Q$ value for a disc of two fluid components:
\begin{align}
    Q_\mathrm{Jog}^2 = \frac{2}{F(\Sigma_1, \Sigma_2, c_1, c_2, k_\mathrm{min})}.
\end{align}

In summary, there is a whole forest of competing $Q$ approximations/formulations, which we also list in Table \ref{tab:existing_definitions}. With the exception of $Q_\mathrm{WS}$, all these $Q$ formulations deliver the correct axisymmetric stability criterion at $Q = 1$. However, disc dynamics focuses on spiral instability and, as noted above, matching axisymmetric stability does not imply that they are good/consistent scales for spiral/swing amplification stability.
Each of them has to be judged by how close to a one-to-one correspondence there is between $Q$ and the maximum achievable swing amplification factor. In the end, this will show the need for a new approximation, which we will provide from Section \ref{sec:proposed_Q_definitions}.  
\begin{table*}
    \centering
    \begin{tabular}{|c|p{75mm}|}
    \hline
        Definition & Notes \\
    \hline
        $Q_\mathrm{Rk} = F_\mathrm{max}(\{\Sigma_{\mathrm{g}, i}\}, \{\Sigma_{\mathrm{s}, j}\}, \{c_i\}, \{\sigma_{R, j}\})^{-1}$ & Best existing definition for a disc of stars and gas \\[1mm]
        $Q_\mathrm{Rf} = F_\mathrm{max}(\{\Sigma_i\}, \{c_i\})^{-1}$ & $Q_\mathrm{Rk}$ if all components are tread as fluid/gas \\[1mm]
        $Q_\mathrm{BR} = F_\mathrm{max}(\Sigma_1, \Sigma_2, c_1, c_2)^{-1}$ & $Q_\mathrm{Rf}$ for a two-component disc (both treated as gas) \\[1mm]
        $Q_\mathrm{RW} = \sum_{i = 1}^N \frac{W_i}{Q_i}$, $W_i = \frac{2 c_m c_i}{c_m^2 + c_i^2}$ & Approximation of $Q_\mathrm{Rf}$; fractional error for N-components $\approx 0.03\sqrt{N}$; $c_m$ is the sound speed of the component with the lowest individual $Q$ \\[1mm]
        $\frac{1}{Q_\mathrm{WS}} = \frac{1}{Q_1} + \frac{1}{Q_2}$ & Q value for a disc treated as two fluids. Erroneous, even for the axisymmetric criterion $Q = 1$ \\[1mm]
        $Q_\mathrm{Jog} = \sqrt{2} F(\Sigma_1, \Sigma_2, c_1, c_2, k_\mathrm{min})^{-1/2}$ & Q value for a disc treated as two fluids. Dispersion relation $s^2$ is minimal at $k = k_\mathrm{min}$. Produces the correct axisymmetric stability criterion at $Q=1$, but no other physical motivation \\
    \hline
    \end{tabular}
    \caption{Existing multi-component definitions of $Q$. Only $Q_\mathrm{Rk}$ is well-motivated for a disc of stars and gas.}
    \label{tab:existing_definitions}
\end{table*}
\section{Local calculations of swing amplification}
To assess swing amplification, we first recur to the standard formulation of shearing sheet instability by
\citet{GoldreichLynden-Bell1965} (GLB1965), who studied linear instabilities to short-wavelength perturbations arising from differential rotation in fluid discs. Following the discussion in \citet{Toomre1981}, we define Cartesian coordinates on a patch of the disc:
\begin{align}
    x &= R - R_0, \\
    y &= R_0 (\phi - \Omega(R_0) t).
\end{align}

\citet{Toomre1981} considered a stellar or fluid (gas) element with guiding centre on $(x, y) = (0, 0)$, and a density wave incident on the star/fluid element at angle $\gamma$ defined by the wavenumbers $k_i$:
\begin{align}
    \tan{\gamma} = \frac{k_x}{k_y} , 
\end{align}
where we demand $k_y < 0$ without loss of generality. The component of the star/fluid element's displacement orthogonal to the wavefront is
\begin{align}
    \xi = x \sin \gamma + y \cos \gamma.
\end{align}
Due to differential disc rotation, the wave shears as
\begin{align}
    \frac{\mathrm{d} (\tan \gamma)}{\mathrm{d} t} = 2 A(R_0),
\end{align}
where $A(R_0)$ is the Oort constant $A(R) = \frac{R}{2} \Omega'(R)$.
Neglecting self-gravity for now, the equation of motion of the star in the rotating coordinate system of the wave is then
\begin{align}
    \ddot{\xi} + \tilde{\kappa}^2(\gamma) \xi &= 0, \\
    \tilde{\kappa}^2(\gamma) &= \kappa^2 - 8 \Omega A\cos^2\gamma + 12 A^2\cos^4\gamma, \label{eq:glb_equation}
\end{align}
where all quantities are calculated at radius $R_0$.  As in \citet{Toomre1981} we refer to equation (\ref{eq:glb_equation}) as the GLB equation, in reference to GLB1965. \cite{Toomre1981} treated the wave instantaneously as a density wave formed by the epicycles of these same star/fluid elements. Assuming the same wave dispersion relation as in the WKB approximation, we replace the element's $\kappa$ by the beat frequency $s$ (see equation \ref{eq:DispersionRelationRk}):
\begin{align}
    \ddot{\xi} + S^2(\gamma)\xi &= 0, \\
    S^2(\gamma) &= s^2(k(\gamma)) \kappa^2 - 8 \Omega A\cos^2\gamma + 12 A^2\cos^4\gamma.\label{eq:combined_sr}
\end{align}
\citet{Toomre1981} and \citet{GoldreichLynden-Bell1965} only considered a single-component stellar and fluid discs respectively. However, their reasoning can be straight-forwardly extended to multi-component discs of stars and gas, using the dispersion relation from \citet{Rafikov2001} in equation (\ref{eq:combined_sr}).

Let us look at the respective stability criteria: 
While $s^2$ (describing axisymmetric instability) is not negative for any value of $k$ in a Toomre-stable disc (with $Q > 1$), $S^2(\gamma)$ (describing swing amplification) can still be negative for some wave inclination angles $\gamma$. $S^2 < 0$ implies that oscillations in $\xi$ will grow (or otherwise shrink). The respective amplification factor for this wave is calculated by integrating $\xi$ as $\gamma$ increases from some minimum to some maximum value. This amplification will depend on the initial (oscillation) phase of $\xi$. We always set the initial phase of $\xi$ such that the amplification is maximal. We sometimes call the result the maximum amplification, though this itself is a function of azimuthal wavenumber (or $m$) which can be maximised.

For discs consisting only of stars we can also use the formalism of \citet{JulianToomre1966} (JT66), who performed a local calculation evolving a density wave in a stellar disc via the collisionless Boltzmann equation (CBE). As explained in \citet{Binney2020}, we can make the Ansatz of a density wave shearing with differential disc rotation:
\begin{align}
    \Sigma_1(\boldsymbol{x}, t) = \tilde{\Sigma}_1 (t) \mathrm{e}^{i \boldsymbol{\mathrm{k}} \cdot \boldsymbol{\mathrm{x}}} ,
\end{align}
where $\boldsymbol{\mathrm{k}}$ is a function of time and rotates such that $\boldsymbol{\mathrm{k}} \cdot \boldsymbol{\mathrm{x}}$ remains constant for any star on a circular orbit. Then the wave's evolution obeys the Julian-Toomre equation (JT equation):
\begin{align}
        \tilde{\Sigma}_1 (t) = \tilde{\Sigma}_{t_\mathrm{i}} (t) + \int_{t_\mathrm{i}}^t \mathrm{d} t' \kappa K(t, t') (\tilde{\Sigma}_\mathrm{e} (t') + \tilde{\Sigma}_1 (t')) , \label{eq:jt_equation}
\end{align}
where $\Sigma_\mathrm{e}(\boldsymbol{\mathrm{x}}, t) = \tilde{\Sigma}_\mathrm{e}(t) \mathrm{e}^{i \boldsymbol{\mathrm{k}} \cdot \boldsymbol{\mathrm{x}}}$ is some imposed external density of the same form as $\Sigma_1$, $\Sigma_{t_i} (\boldsymbol{x}, t) = \tilde{\Sigma}_{t_i}(t)\mathrm{e}^{i \boldsymbol{\mathrm{k}} \cdot \boldsymbol{\mathrm{x}}}$ is the evolution of $\Sigma_1$ from its initial value in the absence of self-gravity and $K(t, t')$ is referred to as the JT kernel. $K(t, t_0)$ is a function of the unperturbed distribution function (DF) $f_0$ (commonly assumed a Schwarzchild distribution with radial velocity dispersion $\sigma_R$). We can extend the JT equation to a disc of $N$ stellar components $i$ with velocity dispersions $\{\sigma_{R, i}\}$:
\begin{align}
    f = f_0 + f_1 = \sum_i f_i = \sum_i f_{i0} + f_{i1}.
\end{align}
For each component $j$ we can treat each perturbation $f_{j1}$ where $j \neq i$ as an external source (ignoring $f_{j0}$ the unperturbed DFs are axisymmetric so $\tilde{\Sigma}_{j0} = 0$). We then have
\begin{align}
    \tilde{\Sigma}_{i1} (t) = \tilde{\Sigma}_{it_i} (t) + \int_{t_i}^t \mathrm{d} t' \kappa K_i(t, t') (\tilde{\Sigma}_e(t') + \sum_{j \neq i} \tilde{\Sigma}_{j1} (t') + \tilde{\Sigma}_{i1} (t')). \label{MultiJTEquation}
\end{align}
Summing over all components $i$ gives the JT equation with kernel
\begin{align}
    K(t, t') = \sum_i K_i(t, t').
\end{align}
To analyse swing amplification \citet{Binney2020} impulsively excited the sheet with an external perturbation $\tilde{\Sigma}_e (t) \propto \delta(t - t_i)$. This created a shearing wave starting with $k = k(t_i)$. The subsequent evolution of $\Sigma_1$ was calculated, as well as its evolution in the absence of self-gravity. The maximum value of $\Sigma_1$ was divided by its maximum value in the absence of self-gravity to give an amplification factor, and the maximum amplification factor was found by varying the time $t_i$ of the impulse, which varies the phase of the wave as in the GLB calculation. 

\begin{figure*}
    \centering
    \includegraphics[width=\textwidth]{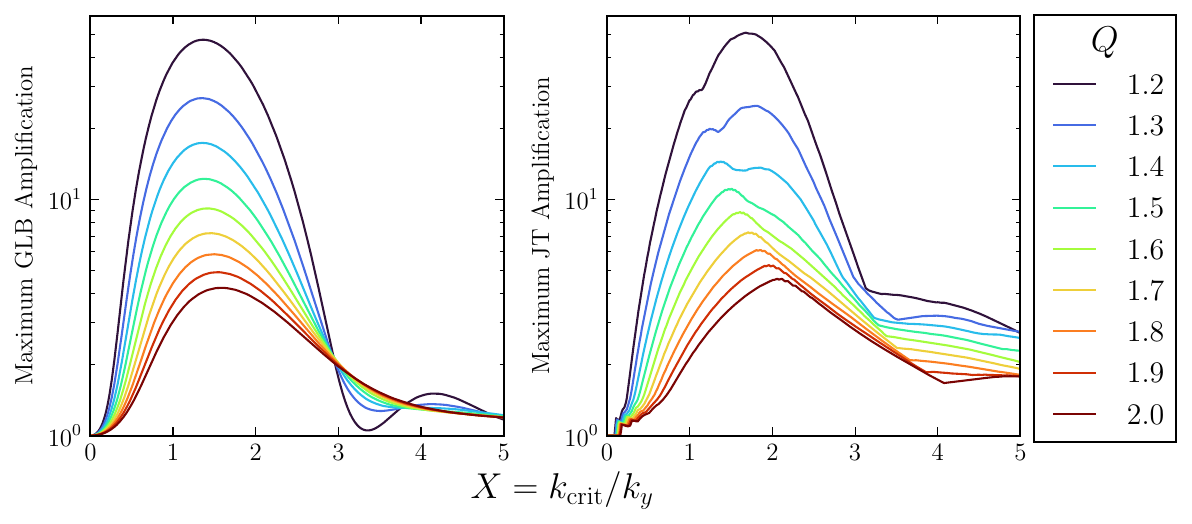}
    \caption{Maximum amplification (choosing optimal initial epicycle phase) achievable in stellar discs of various $Q$ values, as a function of $X = k_\mathrm{crit}/k_y$ (azimuthal wavenumber $k_y = m/R$ for a spiral with multiplicity $m$), as calculated using the Goldreich-Lynden-bell (GLB) equation (equation \ref{eq:glb_equation} and the Julian-Toomre (JT) equation (equation \ref{eq:jt_equation}) respectively. See similar plots in \citet{Toomre1981} and \citet{Binney2020}.}
    \label{fig:glb_jt_compared}
\end{figure*}

Both calculations are local and neglect global effects. The assumption in \citet{JulianToomre1966} and \citet{Binney2020} that the wave shears at the same rate as circular orbits in the disc makes sense in the GLB picture, where the density wave stops oscillating (in the co-rotating frame of a circular orbit) when it is growing ($S^2(\gamma) < 0$), meaning that during amplification the wave shears in lockstep with circular orbits. The behaviour of the wave at other times is oscillatory so does not affect the amplification factor. Otherwise the JT equation is simply an application of the CBE, while the GLB equation as used by \citet{Toomre1981} involves the use of the dispersion relation which is valid in the WKB approximation of tightly wound waves, a condition which is not met when the wave is unwound near $\gamma = 0$. Figure \ref{fig:glb_jt_compared} compares the GLB and JT maximum amplifications for stellar discs of various $Q$ values, against $X$, where
\begin{align}
    X &= \frac{k_\mathrm{crit}}{k_y} \,\, {\mathrm{and}} \,\,
    k_\mathrm{crit} = \frac{\kappa^2}{2 \pi G \Sigma} .
\end{align}
Both approximations show similar behaviour. The JT curve is in places non-differentiable where the global maximum of amplification jumps between two local maxima with respect to initial phase. The GLB equation makes the same assumptions as the JT equation, but in addition makes invalid use of the dispersion relation in the WKB approximation. However, the similarity between the two plots suggests that the GLB equation used in \citet{Toomre1981} is not too far from the truth and might be improved with additional terms. Importantly for us, both approximations display a one-to-one negative correspondence between $Q$ and amplification.  

We now define the maxima (with respect to $X$) from Figure \ref{fig:glb_jt_compared} as the "maximum amplification", and will later plot these against other parameters for discs of various compositions. We sample $250$ values of $X$ up to $X = 5$ and $50$ initial phases for each value of $X$. For the GLB equation we integrate $1000$ timesteps from $\tan \gamma = -16$ to $\tan \gamma = 16$. For the JT equation we calculate the amplification using the code from \citet{Binney2020} with $1000$ timesteps. In addition to the issues with GLB, for stars the JT equation is preferred as the calculation is faster.
\subsection{Evaluating the different \boldmath $Q$ definitions}
To homogenise $Q$ across discs of various compositions, we assert that multi-component discs of given $Q$ should experience the same strength of swing amplification as a single component disc of that same $Q$ value. Here we define a measure to show to what extent each definition deviates from this standard. Consider a two-component disc of stars and gas, and a definition of $Q$ which we wish to evaluate. We define $\tilde{Q} (\Sigma_\mathrm{s}/\Sigma_\mathrm{g}, \sigma_R/c, Q_0)$ to be the $Q$ value (under the given definition) of a disc, with ratios $\Sigma_\mathrm{s}/\Sigma_\mathrm{g}$ and $\sigma_R/c$ given by the first two arguments, which suffers the same maximum amplification under the GLB equation (equation \ref{eq:glb_equation}) as a single-component stellar disc with $Q$ value $Q_0$ (under the given definition). If there is a one-to-one mapping between $Q$ under this definition and maximum GLB amplification, then $\tilde{Q}(\Sigma_\mathrm{s}/\Sigma_\mathrm{g}, \sigma_R/c, Q_0) = Q_0$. We can define a relative error in $Q$ from this ideal:
\begin{align}
    \Delta_\mathrm{GLB} (\Sigma_\mathrm{s}/\Sigma_\mathrm{g}, \sigma_R/c, Q_0) 
    = \frac{\tilde{Q}(\Sigma_\mathrm{s}/\Sigma_\mathrm{g}, \sigma_R/c, Q_0) - Q_0}{Q_0}. \label{eq:delta_glb_definition}
\end{align}
We wish to minimize $|\Delta_\mathrm{GLB}|$ across various values of $\Sigma_\mathrm{s}/\Sigma_\mathrm{g}$, $\sigma_R/c$ and $Q_0$, which is synonymous to approaching a one-to-one mapping between the $Q$ value and amplification under each given definition. For an N-component disc of stars and gas we could analogously define a relative error function $\Delta_\mathrm{GLB}(\{\Sigma_{\mathrm{g}, i}/\Sigma_{\mathrm{g}, 0}\}_{i \neq 0}, \{\Sigma_{\mathrm{s}, j}/\Sigma_{\mathrm{g}, 0}\}, \{c_i/c_0\}_{i \neq 0}, \{\sigma_{R, j}/c_0\}, Q_0)$, though it would be expensive to vary every parameter. Instead, we focus on a realistic distribution: consider a stellar disc with age-velocity dispersion relation (AVR) $\sigma_R(\tau) \propto \tau^\beta$ \citep{Jenkins1992} 
and star formation rate $\mathrm{SFR}(t) \propto \exp(-t/t_\mathrm{SFR})$ (where $t_\mathrm{SFR}$ can be positive or negative). We now treat this many-component stellar disc with the JT equation (equation \ref{eq:jt_equation}) and define
\begin{align}
    \Delta_\mathrm{JT} (t_\mathrm{SFR}, \beta, Q_0) = \frac{\tilde{Q}(t_\mathrm{SFR}, \beta, Q_0) - Q_0}{Q_0} \label{eq:Delta_JT},
\end{align}
where $\tilde{Q}(t_\mathrm{SFR}, \beta, Q_0)$ is the  $Q$ value (under the given definition) of such a disc, with $t_\mathrm{SFR}$ and $\beta$ given by the first two arguments, which suffers the same maximum amplification under the JT equation as a disc with $\beta = 0$ (i.e. a single component disc) of $Q$ value $Q_0$ (under the given definition).
\section{Proposed Definitions of \boldmath $Q$} \label{sec:proposed_Q_definitions}
\subsection{A first, simple approach}
\begin{figure}
    \centering
    \includegraphics[width=0.5\textwidth]{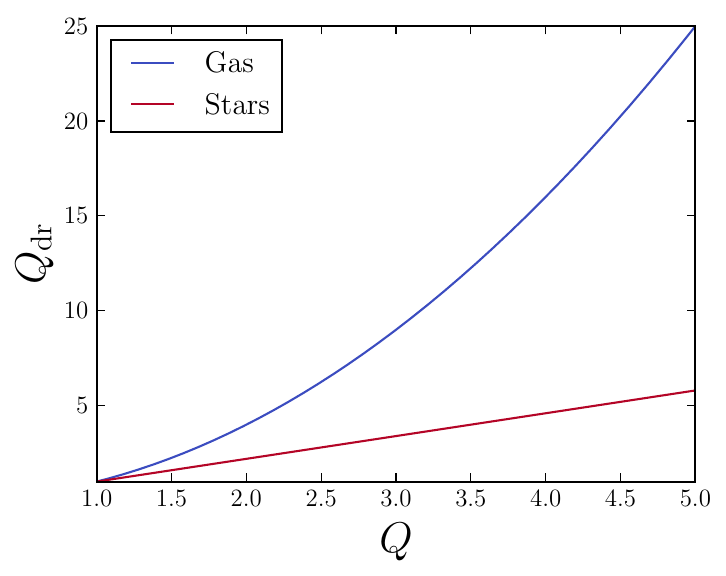}
    \caption{Our first redefinition of $Q$, $Q_\mathrm{dr}$ (equation \ref{eq:Q_dr_definition}), as a function of the respective standard definitions of $Q$ for single component discs of gas or stars. Note that $Q_\mathrm{dr} = Q^2$ for gas, but $Q_\mathrm{dr} \approx Q$ for stars. If $Q_\mathrm{dr}$ is a consistent measure of spiral instability then this implies that the standard definition of $Q$ is inconsistent in comparison between discs of stars and discs of gas when $Q > 1$, and that this inconsistency grows with $Q$.}
    \label{fig:Q_dr}
\end{figure}

We now return to our main problem of describing instability in a mixture of gas and stars.  Inspecting equation (\ref{eq:combined_sr}) we expect that the area above the curve $S^2(\gamma)$ below the $\gamma$ axis, and thus amplification, is maximal when $s^2(k)$ is minimal, so we use this minimum of $s^2(k)$ as predictor of amplification. Formulating the $Q$ value as a function of the dispersion relation minimum,
\begin{align}
    s^2\kappa^2 &= \kappa^2 - 2 \pi G \Sigma |k| + v_s^2 k^2 ,\\ 
    2 v_s^2 |k|_0 &= 2 \pi G \Sigma, \\
    s_{\mathrm{min}}^2 &= 1 - \frac{(\pi G \Sigma)^2}{v_s^2 \kappa^2}, \\
    Q^2 &= \frac{1}{1 - s_\mathrm{min}^2}. \label{eq:Q_gas_squared}
\end{align}
We remind the reader that the task at hand is to find a $Q$ definition that keeps best fixed the relationship between $Q$ and the maximum amplification, which in this subsection we are proposing is a function of the minimum of $s^2(k)$, $s_\mathrm{min}^2$. Therefore we propose to use equation \ref{eq:Q_gas_squared} to define $Q$ both for gas and for stars (as well as for any combination of stellar and gaseous components), where $s_\mathrm{min}^2$ always refers to the minimum of the dispersion relation of the respective medium. We are free to unsquare $Q$ in equation \ref{eq:Q_gas_squared} for our new definition, which preserves the axisymmetric stability criterion at $Q = 1$ and which will turn out to be a more convenient definition when we consider its application to stellar discs. We write the new definition,
\begin{align}
    Q_\mathrm{dr} = \frac{1}{1 - s^2_\mathrm{min}}. \label{eq:Q_dr_definition}
\end{align}
Figure \ref{fig:Q_dr} shows how $Q_\mathrm{dr}$ differs as a function of the standard $Q$ for fluid and stellar discs. While meeting at $Q = Q_{\mathrm{dr}} = 1$, the new $Q_{\mathrm{dr}}$ is quite in line with the stellar disc definition but is the square of the gas $Q$. In other words, if this proves right, the classical fluid $Q$ definition is wrong as a function of $s$ for describing spiral instability.
\begin{figure*}
    \centering
    \includegraphics[width=\textwidth]{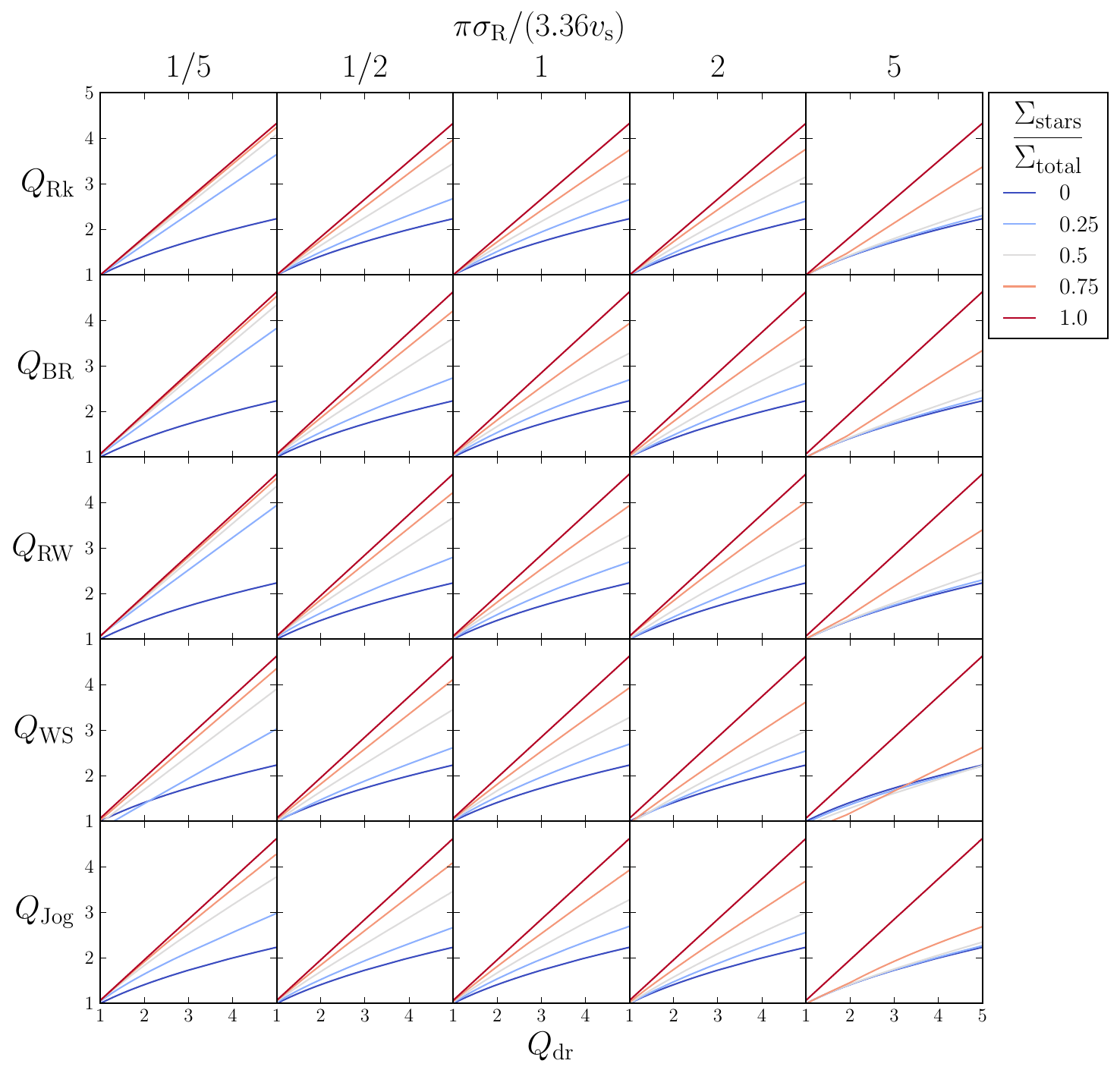}
    \caption{Existing multi-component definitions for $Q$ as a function of our new definition $Q_\mathrm{dr}$ for two-component discs of stars and gas, for various mass ratios and sound-speed/velocity dispersion ratios between the two components. All definitions roughly agree with $Q_\mathrm{dr}$ for a purely stellar disc, while giving a significantly lower value for a gas-only disc. If $Q_\mathrm{dr}$ is a consistent measure of spiral instability then all of these definitions underestimate the correct $Q$ value for $Q > 1$, with the discrepancy growing with $Q$.}
    \label{fig:2c_compared}
\end{figure*}
\begin{figure*}
    \centering
    \includegraphics[width=\textwidth]{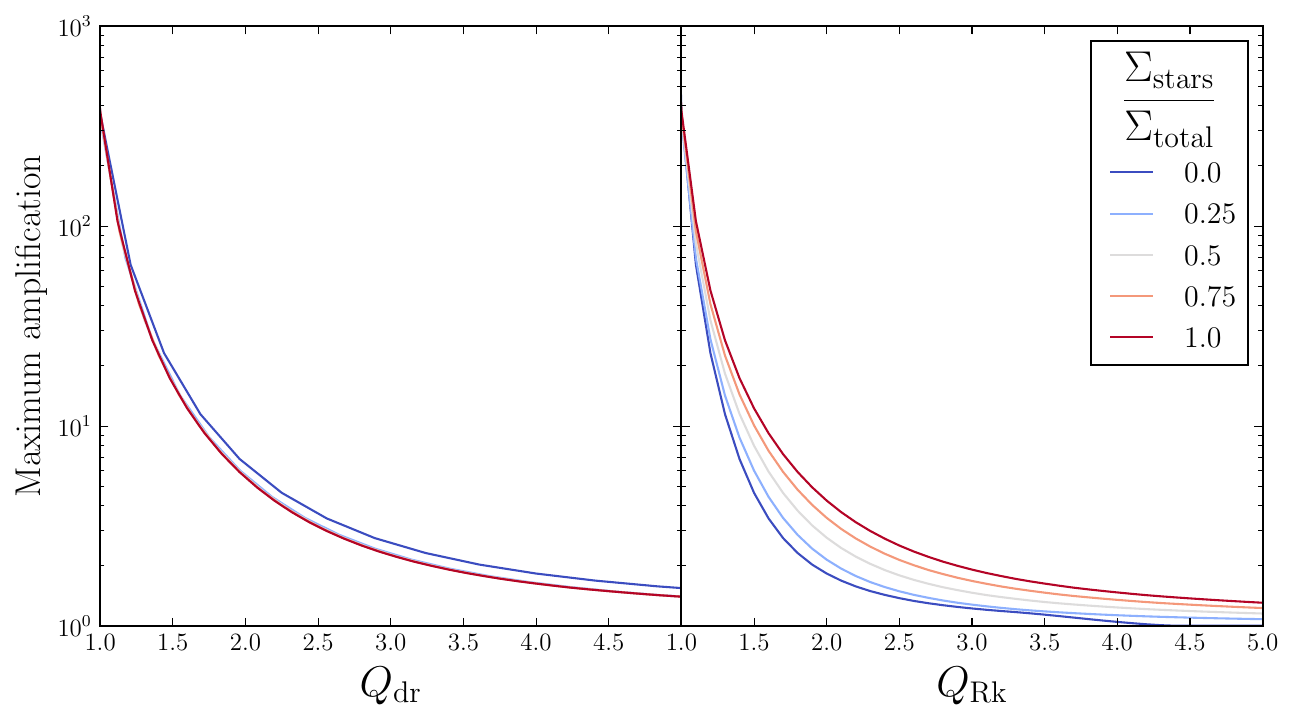}
    \caption{Validation of $Q$ via consistency in amplification:  Maximum amplification (with respect to $X = k_\mathrm{crit} / k_y$ and initial epicycle phase) achievable in a two-component disc of stars and gas with various mass ratios between the two components and with gas sound speed and stellar radial velocity dispersion being almost the same ($c/\pi = \sigma_R/3.36$). Plotted against our simple definition $Q_{\mathrm{dr}}$ (left) vs. the definition from \citet{Rafikov2001}, $Q_{\mathrm{Rk}}$ (right). Note that under the \citet{Rafikov2001} definition, a disc fully consisting of stars has roughly twice the maximum amplification factor as a disc of gas when both have $Q_\mathrm{Rk} = 2.0$. This discrepancy is reduced to about 25\% (in the other direction) when both discs have $Q_\mathrm{dr} = 2.0$}
    \label{fig:2c_amps_dr_rk}
\end{figure*}
We first test this on a two-component disc of stars and gas. Figure \ref{fig:2c_compared} compares the existing multi-component definitions to our definition $Q_{\mathrm{dr}}$ for different surface density ratios (colour) and sound speed/velocity dispersion ratios (columns, written above the top row) between stellar and gas component. We find that all other definitions give systematically lower values for discs with a fluid component than $Q_\mathrm{dr}$ and that this discrepancy is stronger the more the fluid component dominates the instability.

\begin{figure*}
    \centering
    \includegraphics[width=\textwidth]{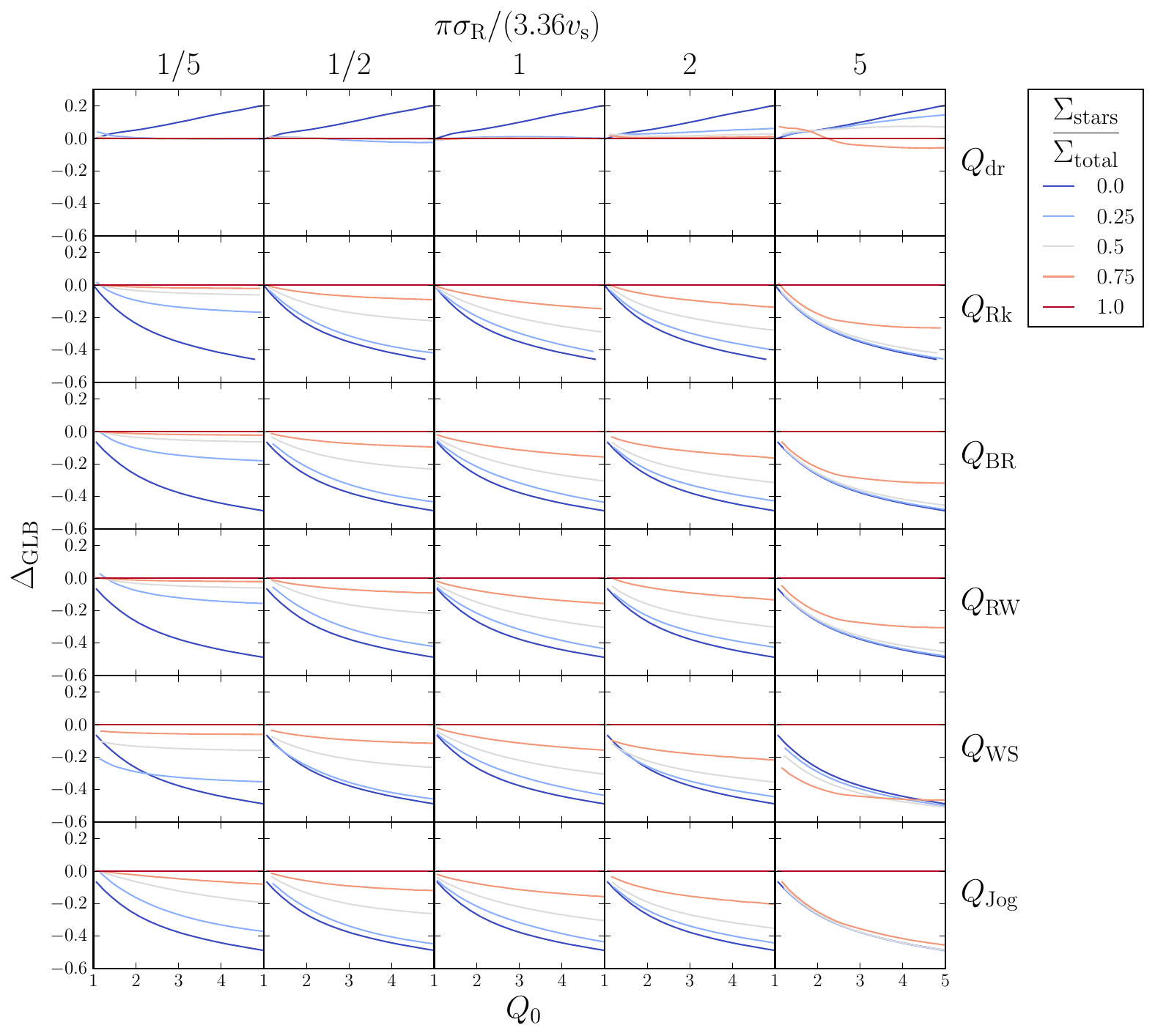}
    \caption{$\Delta_\mathrm{GLB}$ for a 2-component disc of stars and gas as a function of $Q_0$ for $Q_\mathrm{dr}$ and the existing multi-component $Q$ definitions, for various mass ratios and sound speed/velocity dispersion ratios between the two components. Definitions are better, the closer they remain to $0$ across varying $Q_0$. $Q_\mathrm{dr}$ has a closer spread in $\Delta_\mathrm{GLB}$ than any other definition, making $Q_\mathrm{dr}$ closer to one-to-one with amplification than any other definition for 2-component discs of stars and gas. Note that amplification values for each disc are calculated across the domain $1 < Q \leq 5$ for each definition, but that doesn't guarantee that the calculated $\Delta_\mathrm{GLB}$ values fully cover the domain $1 < Q_0 \leq 5$ (see discussion under Figure \ref{fig:delta_JT}).}
    \label{fig:2c_delta_glb}
\end{figure*}

Which ones are better? Figure \ref{fig:2c_amps_dr_rk} shows the maximum swing amplification factor under the GLB equation, plotted against $Q_\mathrm{dr}$ and $Q_\mathrm{Rk}$, for two-component discs with various surface density ratios and $\pi \sigma_R/(3.36 v_\mathrm{s}) = 1$. We recall that (away from $Q = 1$) $Q$ is only sensible/well-defined if the amplification is a unique function of $Q$. It is obvious that $Q_\mathrm{dr}$ has significantly less spread between disc compositions, i.e. is a much better predictor of swing amplification than $Q_\mathrm{Rk}$. To systematically examine this, Figure \ref{fig:2c_delta_glb} plots $\Delta_\mathrm{GLB}$ against $Q_0$ for each definition of $Q$ for a two-component disc of stars and gas. As $Q_0$ increases, the existing definitions give significantly decreasing $Q$ values for discs with a higher gas content, relative to discs with a lower gas content which suffer the same amplification factor under the GLB equation. $Q_\mathrm{dr}$ is preferable with less spread in $\Delta_\mathrm{GLB}$, thus having a map to maximum amplification closer to one-to-one. This map is still not perfect, with gas-only discs in particular still having relatively large $\Delta_\mathrm{GLB}$ as $Q_0$ increases.
\begin{figure*}
    \centering
    \includegraphics[width=\textwidth]{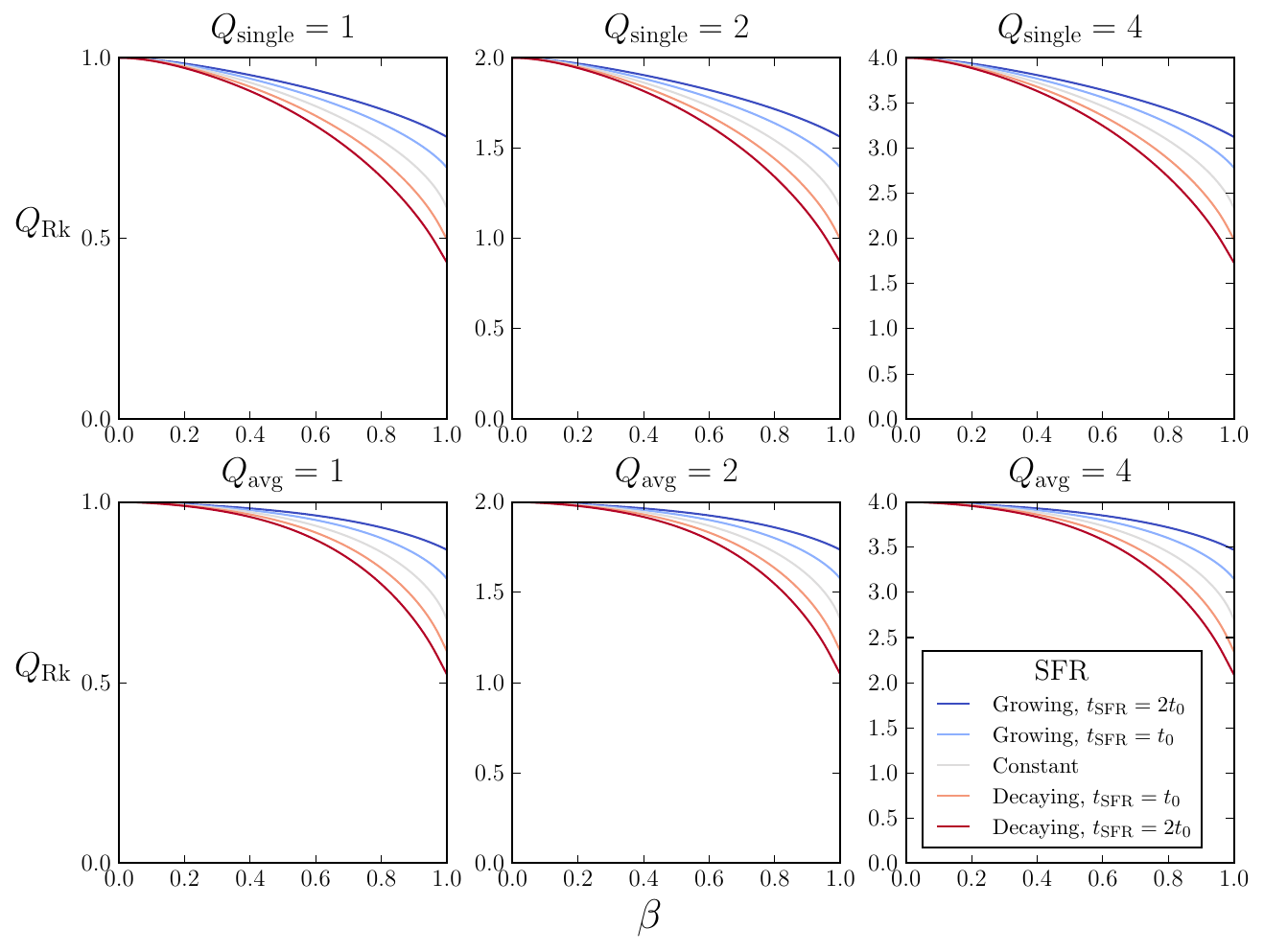}
    \caption{$Q_\mathrm{Rk}$ value of a many-component stellar disc with star formation rate history $\mathrm{SFR}(t) \propto \exp(\pm t/t_\mathrm{SFR})$ and age-velocity dispersion relation $\sigma_R(\tau) \propto \tau^\beta$ as a function of $\beta$, for various SFRs and for various fixed values of $Q$ when the disc is treated as a single component, $Q_\mathrm{single}$, and when the velocity dispersions of the disc are averaged to a single component, $Q_\mathrm{avg}$. Note that both $Q_\mathrm{single}$ and $Q_\mathrm{avg}$ are poor approximations of $Q_\mathrm{Rk}$ for $\beta$ above 0.5, while for the value $\beta = 0.3$ typical for the Solar Neighbourhood \citep{AumerBinney2009, Aumerea2016} $Q_\mathrm{avg}$ might be an acceptable approximation.}
    \label{fig:avr_rk}
\end{figure*}

\begin{figure*}
    \centering
    \includegraphics[width=\textwidth]{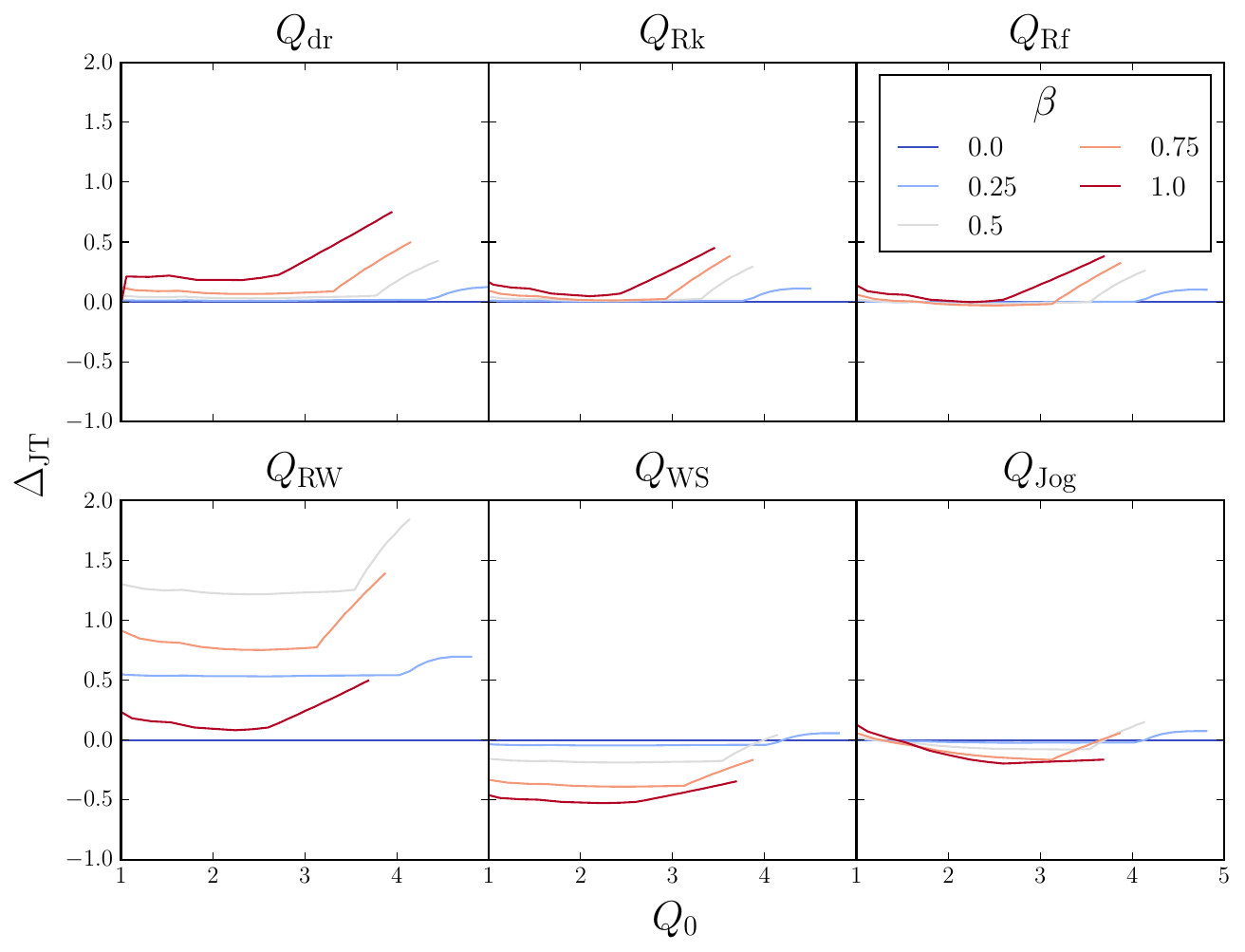}
    \caption{$\Delta_\mathrm{JT}$ (equation \ref{eq:Delta_JT}) as a function of $Q_0$ for $Q_\mathrm{dr}$ and the existing multi-component $Q$ definitions, for a 100-component stellar disc with constant star formation rate (SFR) and with age-velocity dispersion relation (AVR) $\sigma(\tau) \propto \tau^\beta$, for various values of $\beta$. For values of $Q_0 < 3$, $Q_\mathrm{Rk}$ gives the most consistent description of instability across discs of different $\beta$ (while $Q_\mathrm{Rf}$ appears to have a smaller spread in $\Delta_\mathrm{JT}$ in this case, it is poorly physically motivated in comparison to $Q_\mathrm{Rk}$ as the stellar components are treated as fluid (gas) components, so we do not recommend its use for stellar discs). All definitions show diverging $\Delta_\mathrm{JT}$ at $Q_0 \gtrsim 3$; this is explained by, for example, the relatively flat gradients for $Q_\mathrm{dr} \gtrsim 3$ in Figure \ref{fig:avr_amp_dr}, which allow discs of different $\beta$ with greatly differing $Q_\mathrm{dr}$ values to share the same maximum amplification factor. Note that amplification factors are only calculated up to $Q = 5$, so the domain in $Q_0$ is frequently cut off below this value.}
    \label{fig:delta_JT}
\end{figure*}

\begin{figure}
    \centering
    \includegraphics[width=0.5\textwidth]{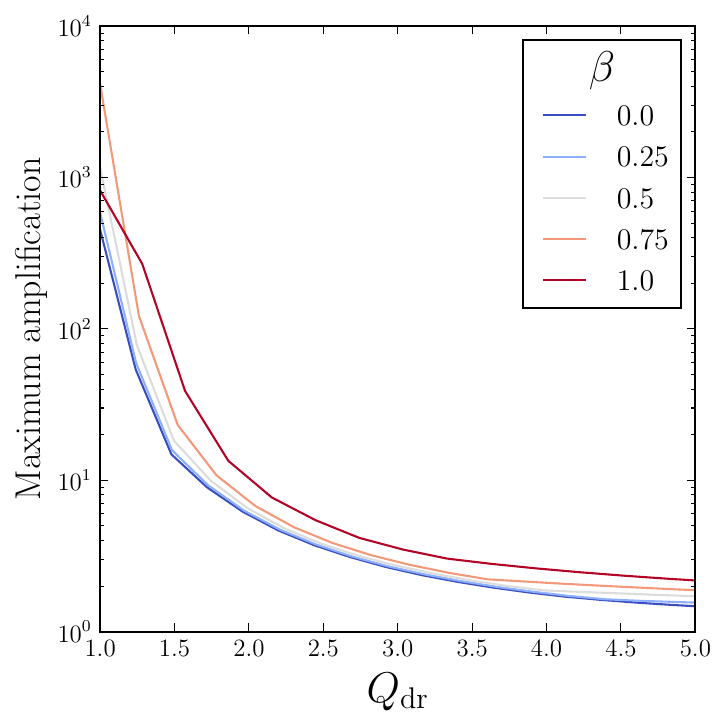}
    \caption{Maximum amplification (with respect to $X = k_\mathrm{crit}/k_y$ and initial epicycle phase) achievable under the Julian-Toomre (JT) equation (equation \ref{eq:jt_equation}) a function of $Q_\mathrm{dr}$ for a many-component stellar disc with constant SFR and AVR $\sigma(\tau) \propto \tau^\beta$, for various values of $\beta$. Note that the relatively low gradients at $Q_\mathrm{dr} \gtrsim 3$ make $Q_\mathrm{dr}$ a poor measure of disc instability above this value, as discs of different $\beta$ which have the same maximum amplification factor have greatly differing $Q$ values.}
    \label{fig:avr_amp_dr}
\end{figure}

We next consider a stellar disc of $100$ components, whose surface densities and velocity dispersions are described by a star formation rate history (SFR) and age-velocity dispersion relation (AVR) respectively. We can demonstrate the need to treat these components separately, rather than treating the stars as a single component and using their overall velocity dispersion, by comparing $Q_\mathrm{Rk}$ with the single-component $Q$ in various regimes. For $\mathrm{SFR}(t) \propto \exp(\pm t/t_\mathrm{SFR})$ and AVR $\sigma_R(t) \propto (t/t_0)^\beta$, the top row of Figure \ref{fig:avr_rk} plots $Q_\mathrm{Rk}$ against $\beta$ for various SFRs, for various fixed values of the $Q$ value of the disc treated as a single component:
\begin{align}
    Q_\mathrm{single} &= \frac{\kappa \sigma_{R,\mathrm{total}}}{3.36 G \Sigma_\mathrm{total}}, \\
    \sigma_{R, \mathrm{total}} &= \sqrt{\frac{1}{\Sigma_\mathrm{total}}\sum_i (\Sigma_i \sigma_{R, i}^2)}, \\
    \Sigma_\mathrm{total} &= \sum_i \Sigma_i.
\end{align}
For a typical $\beta$ value of 0.3 (see \citet{AumerBinney2009, Aumerea2016} and our discussion in Section \ref{sec:solar_neighbourhood}), $Q_\mathrm{Rk}$ is lower than $Q_\mathrm{single}$ by approximately 5\%. Cold components have a destabilising effect which is disproportionate to the mean-square averaging of velocity dispersions. A better approach is take the regular mean of the velocity dispersions. The bottom row of Figure \ref{fig:avr_rk} plots $Q_\mathrm{Rk}$ against $\beta$ for various SFRs and values of $Q_\mathrm{avg}$:
\begin{align}
    Q_\mathrm{avg} &= \frac{\kappa \sigma_{R, \mathrm{avg}}}{3.36 G \Sigma_\mathrm{total}}, \\
    \sigma_{R, \mathrm{avg}} &= \frac{1}{\Sigma_\mathrm{total}} \sum_i(\Sigma_i \sigma_{R, i}).
\end{align}
At $\beta = 0.3$ the discrepancy is reduced and $Q_\mathrm{avg}$ is a reasonable approximation of $Q_\mathrm{Rk}$. For greater accuracy (or for higher values of $\beta$) we must treat the components separately. Figure \ref{fig:delta_JT} plots $\Delta_\mathrm{JT}$, our measure of one-to-oneness between the $Q$ definition and maximum amplification as calculated with the JT equation (see equation \ref{eq:Delta_JT}), against $Q_0$ for each definition of $Q$ for multi-component discs with different age-velocity dispersion relations parametrised by $\beta$ (with a uniform age distribution). For discs with $\beta < 0.5$ and $Q_0 \leq 3$, $Q_\mathrm{Rk}$ gives the most consistent description of JT amplification. The error for $Q_\mathrm{dr}$ is slightly larger, but this is outweighed by the drastically better performance on $\Delta_\mathrm{GLB}$ for discs containing stars and gas. The sudden increase in some curves is explained when we investigate how $\Delta_\mathrm{JT}$ is determined: by comparing fixed amplification factors vs. different $Q$: Figure \ref{fig:avr_amp_dr} shows amplification as a function of $Q_\mathrm{dr}$ for each value of $\beta$. As the curves flatten out at high $Q_\mathrm{dr}$, the vanishing slope implies a diverging measure $\Delta_\mathrm{JT}$ even if the differences in amplification factor remain small compared to their $\beta = 0$ counterparts. We note that  amplifications are only calculated for discs with $Q_\mathrm{Rk}$ between 1 and 5, so most curves terminate at $Q_0 < 5$.
\subsection{A second proposed definition}
So far we have introduced $Q_\mathrm{dr}$: a simple $Q$ instability measure that maps to swing amplification in closer to a single/one-to-one relationship than any other definition and thus improves on the pre-existing $Q_\mathrm{Rk}$ for describing swing amplification in two-component discs of stars and gas. However, while simple to calculate/estimate, it is imperfect/slightly inferior for multi-component stellar discs. In this section we construct a hybrid $Q$ which improves on $Q_\mathrm{Rk}$ for discs containing stellar and gas components while reducing to $Q_\mathrm{Rk}$ in a stars-only system. 

For a single component disc of gas, $Q_\mathrm{dr}$ is the square of the standard $Q$ value and so fulfills a modified form of equation (\ref{eq:1cMarginal}):
\begin{align}
    F_\mathrm{max}(\Sigma, \frac{c}{Q_\mathrm{dr}^{1/2}}) = 1.
\end{align}
For a single component disc of stars the dispersion relation is more complicated (equation (\ref{eq:1c_stellar_dr})). $Q_\mathrm{dr}$ as a function of standard $Q$ is shown by the red line in Figure \ref{fig:Q_dr}, which is close to but not a straight line. Approximating it with the power law $Q_\mathrm{dr} = Q^\alpha$ would give another modified form of equation (\ref{eq:1cMarginal}) for a single component disc of stars:
\begin{align}
    F_\mathrm{max}(\Sigma, \frac{\sigma_R}{Q_\mathrm{dr}^{1/\alpha}}) = 1,
\end{align}
where the exponent $\alpha$ is gained from fitting the aforementioned red line. Let us now abandon $Q_\mathrm{dr}$ and temporarily define $Q_\mathrm{\alpha}$ for a multi-component disc in analogy to the above:
\begin{align}
    F_\mathrm{max}(\{\Sigma_{\mathrm{g}, i}\}, \{\Sigma_{\mathrm{s}, j}\}, \{\frac{c_i}{Q_\alpha^{1/2}}\}, \{\frac{\sigma_{R, j}}{Q_\alpha^{1/\alpha}}\}) = 1.
\end{align}
For a single component stellar disc we would like $Q_\alpha$ to reduce to the standard definition, so that the values are comparable to single-component stellar $Q$ values calculated in the literature. Redefining $Q_\alpha$ as $Q_\alpha^{1/\alpha}$ gives us the definition
\begin{align}
    F_\mathrm{max}(\{\Sigma_{\mathrm{g}, i}\}, \{\Sigma_{\mathrm{s}, j}\}, \{\frac{c_i}{Q_\alpha^{\alpha/2}}\}, \{\frac{\sigma_{R, j}}{Q_\alpha}\}) = 1.
\end{align}
This definition reduces to $Q_\mathrm{Rk}$ in the absence of gas, so inherits its better performance for multi-component stellar discs. However, let us consider further improvement. We reasoned that the lower the minimum of $s^2$ as a function of $k$, the more amplification we can expect in the GLB integration, as $S^2(\gamma)$ will have a deeper/wider trough at negative values. However, the shape of $s^2$ varies and is different for stars and gas. The exponent $\alpha/2$ is derived from the assertion that $Q$ should strictly be a function of $s_\mathrm{min}^2$; we can relax that requirement by introducing a new definition $Q_\mathrm{\delta}$:
\begin{align}
    F_\mathrm{max}(\{\Sigma_{\mathrm{g}, i}\}, \{\Sigma_{\mathrm{s}, j}\}, \{\frac{c_i}{Q_\delta^{1/\delta}}\}, \{\frac{\sigma_{R, j}}{Q_\delta}\}) = 1,
\end{align}
and finding the best $\delta$ to describe swing amplification. This definition is equivalent to
\begin{align}
    F_\mathrm{max} &(\{Q_\delta \Sigma_{\mathrm{g}, i}\}, \{Q_\delta \Sigma_{\mathrm{s}, j}\}, \{Q_\delta^{1 - 1/\delta} c_i\}, \{\sigma_{R, j}\}) = 1, \\
    \frac{1}{Q_\delta} &= F_\mathrm{max} (\{\Sigma_{\mathrm{g}, i}\}, \{\Sigma_{\mathrm{s}, j}\}, \{Q_\delta^{1 - 1/\delta} c_i\}, \{\sigma_{R, j}\})  . \label{eq:Q_delta_definition}
\end{align}
Physically interpreted in the context of equation (\ref{eq:Q_BR_multiplied_densities}) and \citet{BertinRomeo1988}, $Q_\delta$ is defined such that we obtain a disc of marginal axisymmetric stability if we multiply the stellar and gas surface densities of our disc by  $Q_\mathrm{\delta}$ (at fixed $\sigma_i$ for the stellar components), and multiply the gas component sound speeds by $Q_\mathrm{\delta}^{1 - 1/\delta}$ to account for the higher order scaling of instability in fluid components. We can iteratively calculate the self-consistent solution for $Q_\delta$:
\begin{align}
    \frac{1}{Q_{\delta, n}} &= F_\mathrm{max} (\{\Sigma_{\mathrm{g}, i}\}, \{\Sigma_{\mathrm{s}, j}\}, \{Q_{\delta, n - 1}^{1 - 1/\delta} c_i\}, \{\sigma_{R, j}\}).
\end{align}
If we start with $Q_{\delta, 0}$ = 1, we get $Q_{\delta, 1} = Q_\mathrm{Rk}$, and in the following steps $Q_{\delta, n}$ converges monotonically onto $Q_\delta$. 

\begin{figure*}
    \centering
    \includegraphics[width=\textwidth]{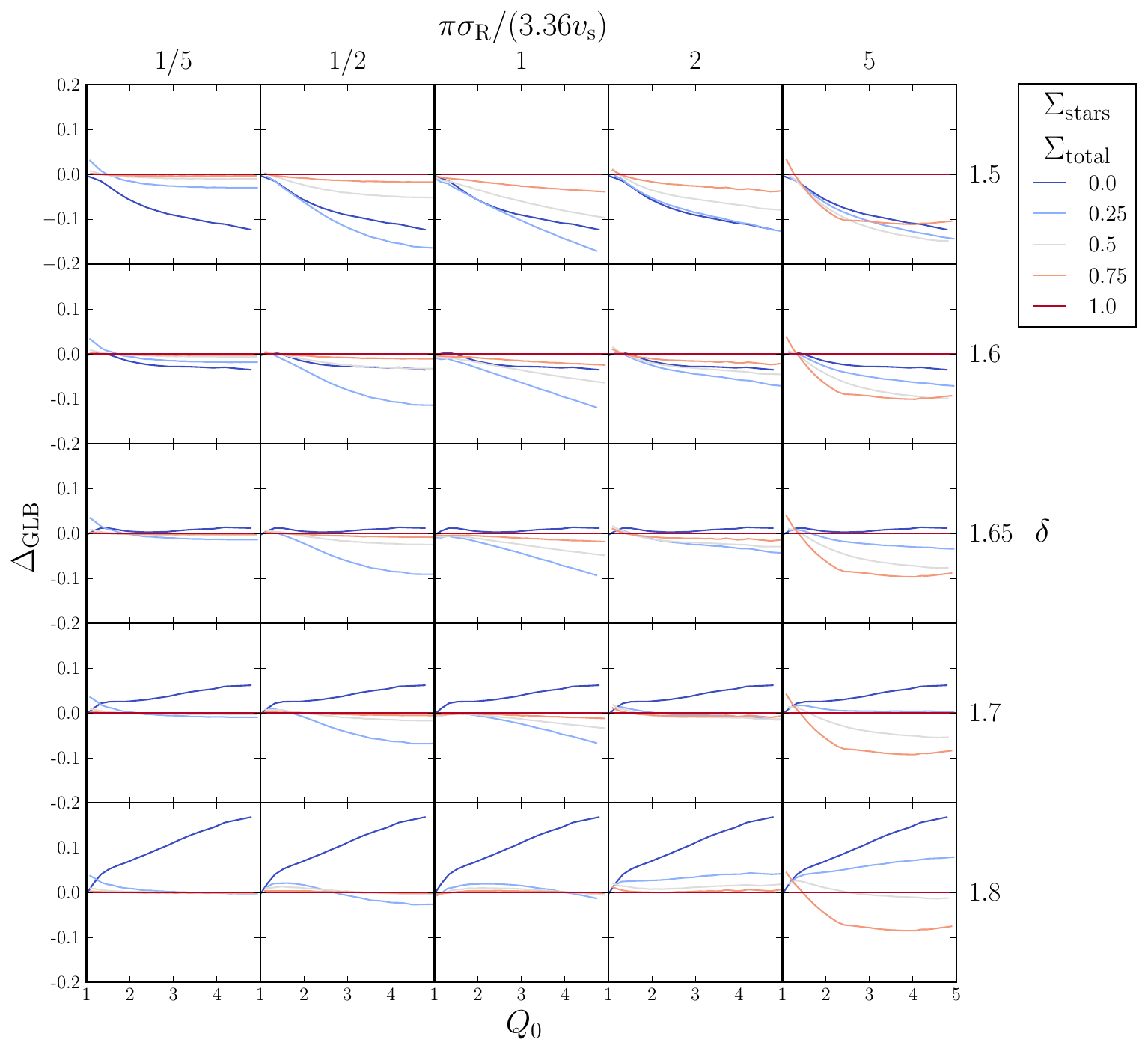}
    \caption{$\Delta_\mathrm{GLB}$ as a function of $Q_0$ for $Q_\mathrm{\delta}$ with various values of $\delta$, for two-component discs of stars and gas with various mass ratios and sound-speed/velocity dispersion ratios between the two components. The $\Delta_\mathrm{GLB}$ values seem to be most consistent and overall closest to zero for $\delta \approx 5/3$, so we adopt $Q_{\delta = 5/3}$ as our preferred definition of $Q$ for thin, multi-component discs. In comparison with Figure \ref{fig:2c_delta_glb}, $Q_{\delta=5/3}$ is an improvement on $Q_\mathrm{dr}$ and all existing definitions. Note that amplification values for each disc are calculated across the domain $1 < Q_\delta \leq 5$ for each value of $\delta$, but that doesn't guarantee that the calculated $\Delta_\mathrm{GLB}$ values fully cover the domain $1 < Q_0 \leq 5$ (see discussion under Figure \ref{fig:delta_JT}).}
    \label{fig:2c_delta_glb_hybrid}
\end{figure*}

\begin{figure*}
    \centering
    \includegraphics[width=\textwidth]{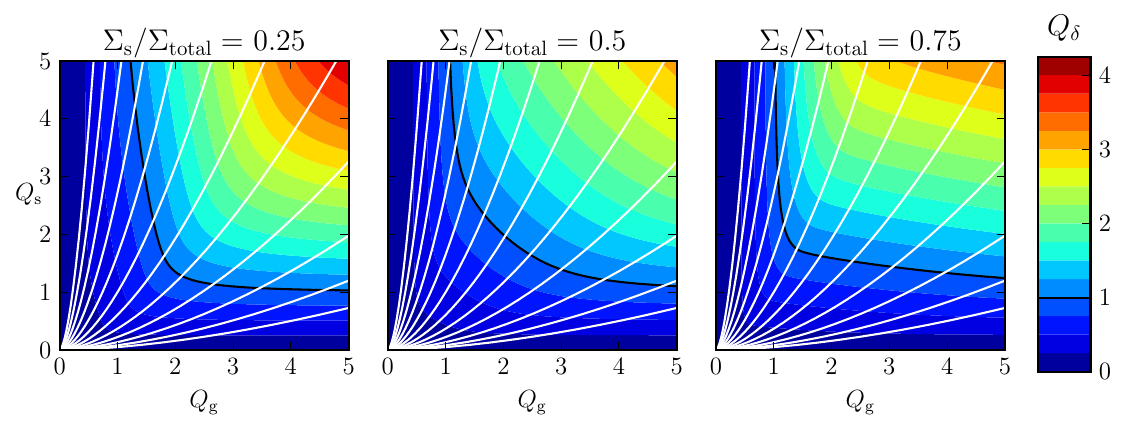}
    \caption{Contours of constant $Q_{\delta = 5/3}$ for two-component discs of stars and gas with various mass ratios between the components, as a function of the single-component $Q$ values of the gas and stellar components, $Q_\mathrm{g}$ and $Q_\mathrm{s}$. The axisymmetric stability criterion is preserved as $Q_\delta = 1$, so the black $Q_\delta = 1$ contours are the same as in Figure \ref{fig:q_rk_contours}. Each contour is an enlargement of the $Q_\delta = 1$ contour along the lines $Q_\mathrm{s} \propto Q_\mathrm{g}^{5/3}$ (in white), by factor $Q_\mathrm{\delta}$ in the $Q_\mathrm{s}$ axis, whereas the $Q_\mathrm{Rk}$ contours in Figure \ref{fig:q_rk_contours} are simple linear enlargements of the $Q_\mathrm{Rk} = 1$ contour.}
    \label{fig:q_hybrid_contours}
\end{figure*}

Figure \ref{fig:2c_delta_glb_hybrid} shows $\Delta_\mathrm{GLB}$ for two-component discs of stars and gas with various values of $\Sigma_\mathrm{s}/\Sigma_\mathrm{total}$ and $\sigma_R/c$, for $Q_\delta$ with multiple values of $\delta$. $\delta \approx 5/3$ appears to give the closest to a one-to-one correspondence between $Q_\delta$ and maximum amplification. Figure \ref{fig:q_hybrid_contours} shows $Q_{\delta=5/3}$ plotted as a function of $Q_\mathrm{g}$ and $Q_\mathrm{s}$ for different values of $\Sigma_\mathrm{s}/\Sigma_\mathrm{total}$. Note that rather than being linear enlargements of the $Q_\delta = 1$ contour from the origin, the contours in $Q_\delta$ are now enlargements of the $Q_\delta = 1$ contour, by factor $Q_\delta$ in the $Q_\mathrm{s}$ axis, along the lines $Q_\mathrm{g} \propto Q_\mathrm{s}^{3/5}$.
\begin{figure*}
    \centering
    \includegraphics[width=\textwidth]{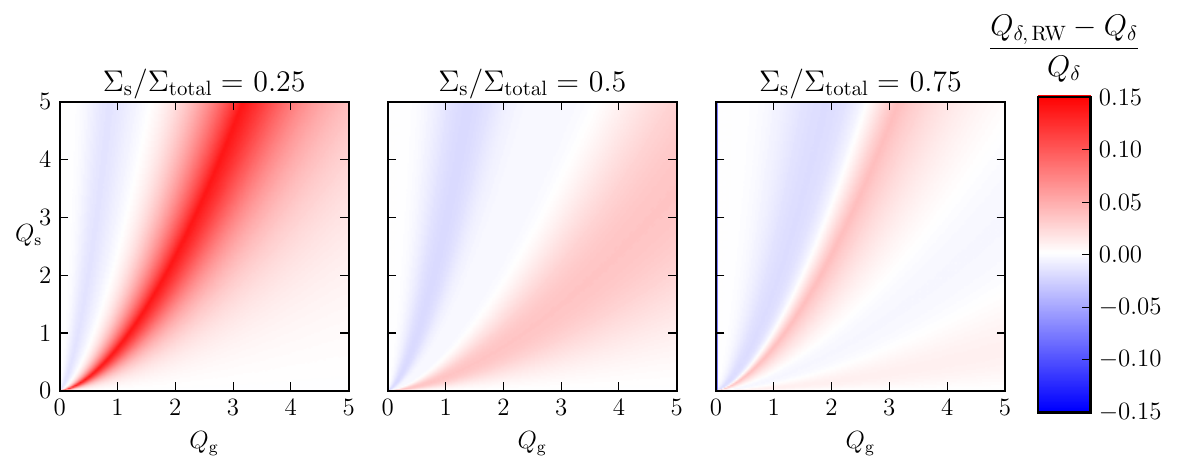}
    \caption{Error in $Q_{\delta = 5/3, \mathrm{RW}}$ as an approximation of $Q_{\delta = 5/3}$, for 2-component discs of stars and gas with various mass ratios between the two components, as a function of the single-component $Q$ values of the gas and stellar components, $Q_\mathrm{g}$ and $Q_\mathrm{s}$. Note: at $Q_\mathrm{g} \approx 0$ the error goes much lower than -0.15, so we cut off the color scale at -0.15 to keep the rest of the plot readable.}
    \label{fig:q_hybrid_rw_error}
\end{figure*}

As discussed in Section \ref{sec:existing_definitions}, $Q_{\mathrm{RW}}$ has inferior accuracy to $Q_{\mathrm{Rk}}$. Since it is somewhat more straight-forward to calculate, however, we provide an analogous iterative extension: 
\begin{align}
    Q_{\delta, \, \mathrm{RW}, n} &= 
    \begin{cases}
        \frac{1}{Q_{\delta, \mathrm{\, RW}, n - 1}^{1 - 1/\delta} Q_\mathrm{g}} + \frac{W_n}{Q_{\delta, \mathrm{ \, RW}, n - 1} Q_\mathrm{s}} & Q_\mathrm{s} > Q_\mathrm{g}\\
        \frac{W_n}{Q_{\delta, \mathrm{ \, RW}, n - 1}^{1 - 1/\delta} Q_\mathrm{g}} + \frac{1}{Q_{\delta, \mathrm{ \, RW}, n - 1} Q_\mathrm{s}} & Q_\mathrm{s} < Q_\mathrm{g}
    \end{cases}, \\ 
    W_n &= \frac{2 Q_{\delta, \mathrm{ \, RW}, n - 1}^{1 - 1/\delta} \frac{c \sigma_R}{3.36 \pi}}{(Q_{\delta, \mathrm{ \, RW}, n - 1}^{1 - 1/\delta} \frac{c}{\pi})^2 + (\frac{\sigma_R}{3.36})^2} , 
\end{align}
where again it is beneficial to start the series with $Q_{\delta, \mathrm{ \, RW}, 0} = 1$, allowing for monotonic convergence. Figure \ref{fig:q_hybrid_rw_error} shows the fractional error of $Q_{\delta,\, \mathrm{RW}}$ on $Q_\delta$ as a function of $Q_\mathrm{g}$ and $Q_\mathrm{s}$ for various $\Sigma_\mathrm{s}/\Sigma_\mathrm{total}$ values. We see that $Q_{\delta, \, \mathrm{RW}}$ is a good approximation for $Q_\mathrm{\delta}$ for two-component discs which are not gas-dominated. We expect the error to scale as $\sqrt{N}$ with the number of components $N$ as is the case for $Q_\mathrm{RW}$ as an approximation of $Q_\mathrm{Rf}$. We recommend $Q_{\delta, \mathrm{RW}}$ as an approximation for $Q_\delta$ only in cases where computing resources are extremely limited.
\subsection{Realistically thick discs}
We will now apply a similar treatment to realistically thick discs, using existing finite-thickness corrections to the WKB dispersion relation. These corrections are based on problematic assumptions, so we present this treatment as a curiosity but would not strongly recommend its application at this stage. \citet{Vandervoort1970a} calculated an equilibrium DF for a highly flattened disc of finite thickness using the epicycle approximation (near-circular orbits), assuming a large vertical vs radial oscillation frequency for stars. \citet{Vandervoort1970b} studied density waves as a perturbation on this DF, and calculated an approximate dispersion relation. When $k \langle z \rangle < 2$, where $\langle z \rangle$ is the scale height of the disc, the dispersion relation is well approximated to first order in $\langle z \rangle/R$ by multiplying $\Sigma$ in the dispersion relation for an infinitesimally thin disc (equation \ref{eq:1c_stellar_dr}) by the reduction factor:
\begin{align}
    \mathcal{I} = \frac{1}{1 - k \langle z\rangle}.
\end{align}

\citet{Romeo1992} argued that this result could be extended to a two-component disc of stars and gas, provided the two components are not "strongly coupled". This demands that stars and gas have active modes in disjoint regions of wavenumber $k$, while our results show an overlap (see Section \ref{sec:solar_neighbourhood}, Figure \ref{fig:mw_dr_polarisations}). It is also problematic if one tries to model a composite disc with a range of velocity dispersions (e.g. age-dispersion relation), where neighbours cover very similar $k$ ranges. \citet{Romeo2013} argued, however, that finite-thickness effects could be estimated in this way for a 10-component disc. This \citet{Romeo1992} two-component extension proceeds as follows: For each component (with surface density $\Sigma$, dispersion/sound speed $c$, $\sigma_z$) they defined an exponential scale height, if this component were isothermal in isolation:
\begin{align}
    z_{\mathrm{E, \, g}} &= \frac{c^2}{2 \pi G \Sigma_\mathrm{g}}, \\
    z_{\mathrm{E, \, s}} &= \frac{\sigma_z^2}{2 \pi G \Sigma_\mathrm{s}}.
\end{align}
They then obtained the effective scale heights $z_{\mathrm{eff}, \, i} = \Sigma_i/(2 \rho_{0i})$ where $\rho_{0i}$ is the central volume density of component $i$:
\begin{align}
    z_{\mathrm{eff}, \, s} &= 2 z_\mathrm{E, \, s} \left(\frac{1}{2} \sqrt{1 + \gamma \beta_z} \int_0^1 \frac{\mathrm{d} u}{\sqrt{(1 - u) + \gamma \beta_z (1 - u^{\beta_z^{-1}})}}\right), \\
    z_{\mathrm{eff}, \, g} &= 2 z_\mathrm{E, \, g} \left(\frac{1}{2} \sqrt{1 + \gamma \beta_z} \int_0^1 \frac{\mathrm{d} v}{\sqrt{(1 - v^{\beta_z}) + \gamma \beta_z (1 - v)}}\right),
\end{align}
where $\beta_z = c^2/(\sigma_z^2)$ and $\gamma =\rho_{0 \mathrm{g}}/(\rho_{0 \mathrm{s}})$ is well approximated by $\Sigma_\mathrm{g}/(\Sigma_\mathrm{s}) = \gamma \beta_z^{3/5}$ when $0 < \gamma \leq 1$ and $0 < \beta_z \leq 1$, as discussed in \citet{Romeo1992}. They then multiplied each $\Sigma_i$ by the reduction factor
\begin{align}
    \mathcal{I}_i = \frac{1}{1 - k z_{\mathrm{eff}, i}}.
\end{align}
In the formalism we adopted here this yields the axisymmetric stability criterion:
\begin{align}
    F_\mathrm{max} (\Sigma_\mathrm{g}, \Sigma_\mathrm{s}, c, \sigma_R, \sigma_z) < 1.
\end{align}
where $F (\Sigma_\mathrm{g}, \Sigma_\mathrm{s}, c, \sigma_R, \sigma_z)$ = $F (\mathcal{I}_\mathrm{g} \Sigma_\mathrm{g}, \mathcal{I}_\mathrm{s} \Sigma_\mathrm{s}, c, \sigma_R)$ as defined in equation (\ref{eq:F_definition}). $F$ (as well as the WKB dispersion relation) is invariant under the map:
\begin{align}
\begin{aligned}
    \Sigma_\mathrm{g} \rightarrow A \Sigma_\mathrm{g} \mathrm{;}& \,\, \Sigma_\mathrm{s} \rightarrow A \Sigma_\mathrm{s} \\
    \sigma_R \rightarrow A \sigma_R \mathrm{;}& \,\,
    \sigma_z \rightarrow A \sigma_z \\
    k \rightarrow {k}/{A} \mathrm{;}& \,\,    c \rightarrow A c .
\end{aligned}
\end{align}
\citet{Romeo1992} suggested, for observational convenience, to parametrise our own Galaxy in terms of $\delta_\mathrm{s} = \sigma_z/\sigma_R$ ($\delta_\mathrm{g} = 1$ by isotropy of a collisional fluid), and external galaxies by $z_\mathrm{eff, \, s}$ and $z_\mathrm{eff, \, g}$. These two parametrisations affect how we might define a $Q$ value analogously to \citet{BertinRomeo1988}. Dividing the individual $Q$ values of all components by $Q_\mathrm{Rk}$ in an infinitesimally thin disc gives a disc of marginal stability. In a thick disc we introduce extra parameters, so in the two-component case our extension of $Q_\mathrm{Rk}$ is defined
\begin{align}
    F_\mathrm{max}(A\Sigma_\mathrm{g}, A\Sigma_\mathrm{s}, Bc, B\sigma_R, C\sigma_z) = 1.
\end{align}
where $A/B = Q_{\mathrm{Rk}, z}$, and the definition of $C$ depends on what we hold constant. If we parametrise the disc by $\delta_\mathrm{s}$, we will compare the disc to a disc of the same $\delta_\mathrm{s}$ which is marginally stable to axisymmetric perturbations, so can define $Q_\mathrm{Rk, z}$:
\begin{align}
    F_\mathrm{max}(\Sigma_\mathrm{g}, \Sigma_\mathrm{s}, \frac{c}{Q_{\mathrm{Rk, }z}}, \frac{\sigma_R}{Q_{\mathrm{Rk}, z}}, \frac{\sigma_z}{Q_{\mathrm{Rk}, z}}) = 1, \label{eq:delta_s_constant}
\end{align}
as we must divide $\sigma_z$ by the same value as $\sigma_R$ to hold $\delta_\mathrm{s}$ constant. If we parametrise the disc by $z_\mathrm{eff, \, g}$ and $z_\mathrm{eff, \, s}$, then to hold each constant we must hold $\beta_z = c^2/\sigma_z^2$ and $\Sigma_\mathrm{g}/\Sigma_\mathrm{s}$ constant, as well as $z_\mathrm{E, g} \propto c^2/\Sigma_g$ and $z_\mathrm{E, s} \propto \sigma_z^2/\Sigma_s$. These constraints are satisfied by the definition
\begin{align}
    F_\mathrm{max}(Q_{\mathrm{Rk}, z}^2 \Sigma_\mathrm{g}, Q_{\mathrm{Rk}, z}^2 \Sigma_\mathrm{s}, Q_{\mathrm{Rk}, z} c, Q_{\mathrm{Rk}, z} \sigma_R, Q_{\mathrm{Rk}, z} \sigma_z) = 1,
\end{align}
which under the invariance of $F_\mathrm{max}$ to scaling of its arguments is equivalent to equation (\ref{eq:delta_s_constant}). Therefore $Q_{\mathrm{Rk}, z}$ has the same definition regardless of whether we fix $\delta_\mathrm{s}$ or effective scale heights as the characteristic parameter for vertical extent.

Instead of holding a particular parameter constant, we now test different definitions against the GLB equation (combining WKB with the thickness reduction factors) as previously. We define $Q_{\delta, \epsilon}$:
\begin{align}
    F_\mathrm{max} (\Sigma_\mathrm{g}, \Sigma_\mathrm{s}, \frac{c}{Q_{\delta, \epsilon}^{1/\delta}}, \frac{\sigma_R}{Q_{\delta, \epsilon}}, \frac{\sigma_z}{Q_{\delta, \epsilon}^{1/\epsilon}}) = 1. \label{eq:q_delta_epsilon_definition}
\end{align}
We find $\Delta_\mathrm{GLB}$ to be optimal around $\delta = 5/3$ and $\epsilon$ between $\frac{11}{6}$ and 2 (See figures \ref{fig:epsilon=11/6} and \ref{fig:epsilon=2}, both of which show maximum $\Delta_\mathrm{GLB}$ values of roughly 0.1 for a realistic range of disc parameters). This means that the optimal $Q_{\delta, \epsilon}$ is approximately defined:
\begin{align}
    F_\mathrm{max} (Q_{\delta, \epsilon} \Sigma_\mathrm{g}, Q_{\delta, \epsilon} \Sigma_\mathrm{s}, Q_{\delta, \epsilon}^{2/5} c, \sigma_R, Q_{\delta, \epsilon}^{1/2} \sigma_z) = 1.
\end{align}
The optimal $Q_{\delta, \epsilon}$ is therefore physically interpreted as the value by which, to obtain a disc marginally stable to axisymmetric perturbations, you would multiply the surface densities in the disc while keeping the stellar radial velocity dispersion constant, multiplying the gas sound speed by $Q_{\delta, \epsilon}^{2/5}$ to account for the higher order scaling of instability in fluids, and keeping the effective thicknesses $z_{\mathrm{eff}, s}$ and $z_{\mathrm{eff}, g}$ of the disc approximately constant ($\beta_z \rightarrow \beta_z/Q_{\delta, \epsilon}^{1/5}$ and $z_\mathrm{E, g} \rightarrow z_\mathrm{E, g}/Q_{\delta, \epsilon}^{1/5}$, so the changes in $z_{\mathrm{eff}, s}$ and $z_{\mathrm{eff}, g}$ are negligible for $Q_{\delta, \epsilon}$ sufficiently close to 1). This physical interpretation puts the definition $Q_{\delta=5/3, \epsilon=2}$ in close analogy with the definition of $Q_{\delta = 5/3}$ for a thin disc: rather than ignoring thickness we just hold it approximately constant.

We find $Q_{\delta = 5/3, \epsilon \approx 2}$ best describes the stability of a thick, 2-component disc to spiral instabilities. As well as being complicated to calculate it is based on problematic assumptions, so would need to be validated against simulations if it were to ever be used.
\section{Applications}
\subsection {Application to the Solar Neighbourhood} \label{sec:solar_neighbourhood}
\begin{figure}
    \centering
    \includegraphics[width=\linewidth]{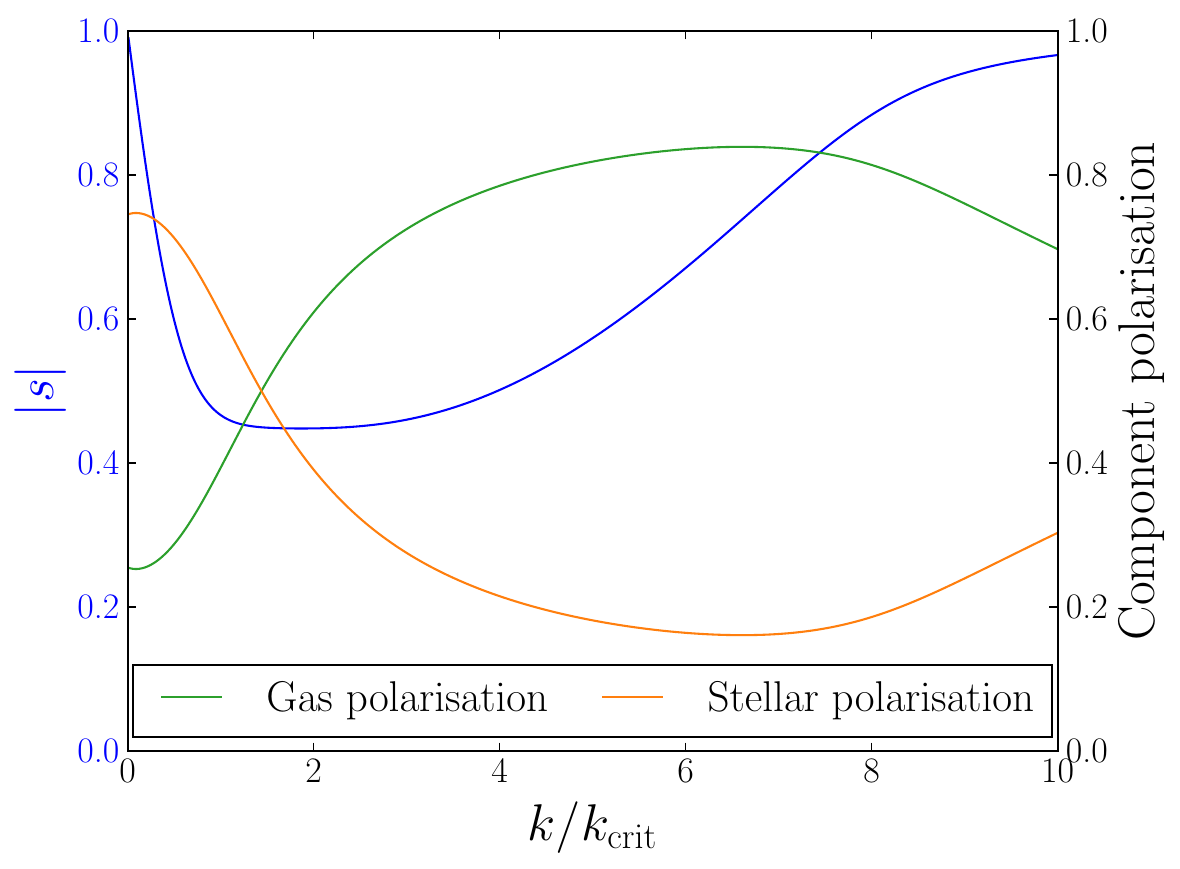}
    \caption{LSK dispersion relation \citep{Rafikov2001} of the Galactic disc at the Solar Neighbourhood, using the gas properties in Table \ref{tab:gas_properties} \citep{HolmbergFlynn2000}, the stellar properties listed in Section \ref{sec:solar_neighbourhood} from \citet{AumerBinney2009,  BovyRix2013} and the epicycle frequency $\kappa = 41 \mathrm{\,km\,s}^{-1}\mathrm{kpc}^{-1}$ from \citet{Bland-hawthornGerhard2016}. Also plotted are the total polarisation functions (relating to the fractional contribution of each component to the pattern) of the stellar and gas components; the dispersion relation is calculated such that the sum of the polarisations is 1 at all $k$.}
    \label{fig:mw_dr_polarisations}
\end{figure}

\begin{figure*}
    \centering
    \includegraphics[width=\textwidth]{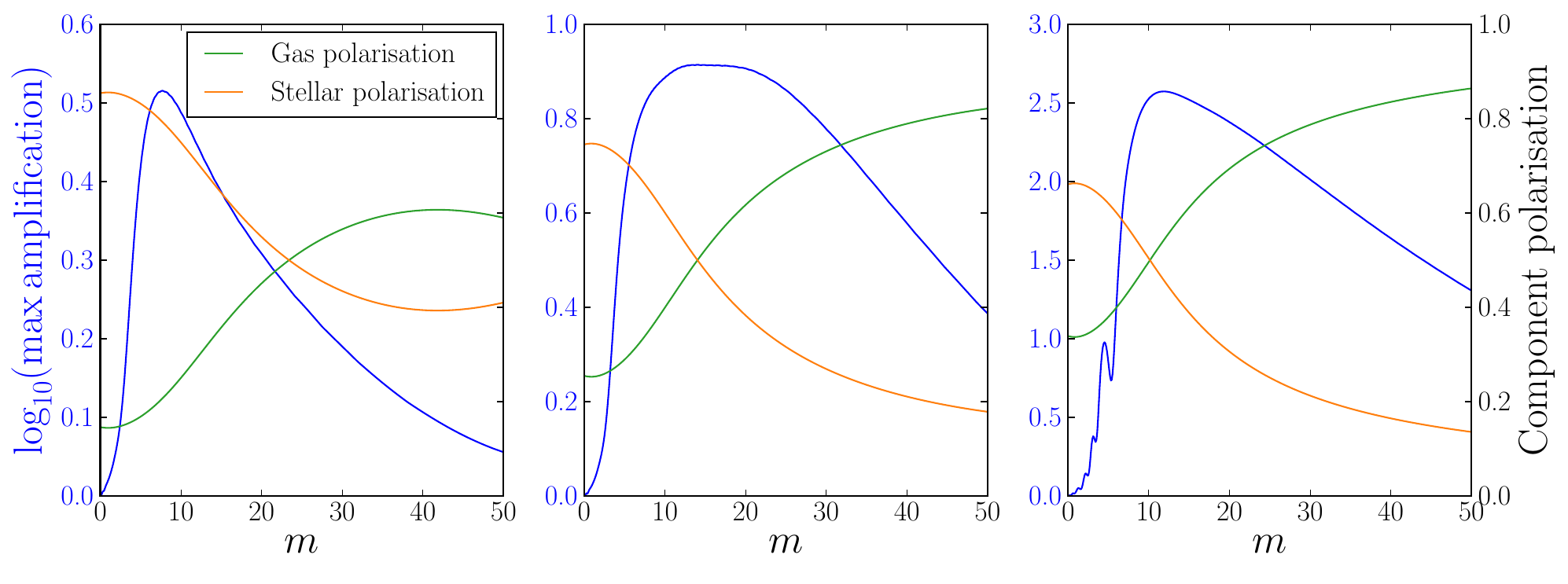}
    \caption{Maximum amplification under the GLB equation, given the properties of the Solar Neighbourhood stated in Section \ref{sec:solar_neighbourhood} (including Table \ref{tab:gas_properties}), as a function of $m$. The total surface density of the gas is at its lower bound in the left panel, central value in the middle panel and upper bound in the right panel. The associated total polarisations for gas and stars (equations \ref{eq:gas_polarisation} and \ref{eq:stellar_polarisation}) are calculated at $k = \sqrt{3/2} k_\mathrm{crit}/X$: the wavenumber when the reduced spring rate $\tilde{\kappa}^2$ as described in \citet{Toomre1981} is at its lowest value. It appears that lower-multiplicity spiral structure would be dominated by stellar contributions, while higher-multiplicity structure would be gas-dominated. Note that maximum amplification in all cases occurs at $m \approx 10$, suggesting that lower $m$-numbers are preferred by the feedback mechanism of swing amplification to give the Milky way an $m \approx 4$ spiral pattern despite that $m$-number not having the strongest amplification.}
    \label{fig:m_amplification_errors}
\end{figure*}

\begin{table}
    \centering
    \begin{tabular}{|c|c|c|}
    \hline
        Description & $\Sigma_i/\mathrm{M}_\odot\mathrm{pc}^{-2}$ & $c_i/\mathrm{km\,s}^{-1}$ \\
    \hline
        $\mathrm{H}_2$ & 3.0 & 4.0 \\
        $\mathrm{H}_\mathrm{I}$(1) & 4.0 & 7.0 \\
        $\mathrm{H}_\mathrm{I}$(2) & 4.0 & 9.0 \\
        warm gas & 2.0 & 40.0 \\
        \hline
    \end{tabular}
    \caption{Properties of the ISM in the Solar Neighbourhood \citep{HolmbergFlynn2000}. The total surface density has an uncertainty of about 50$\%$.}
    \label{tab:gas_properties}
\end{table}
We now apply our methods to the Solar Neighbourhood. \citet{HolmbergFlynn2000} gave properties for the  multi-phase ISM in the Solar Neighbourhood (see Table \ref{tab:gas_properties}). Note that the surface density of the gas has an uncertainty of about 50$\%$. \citet{Bland-hawthornGerhard2016} reviewed measurements of the Galactic rotation curve in the literature, finding a circular frequency for the Sun of $29.0 \pm 1.8 \mathrm{\,km\,s}^{-1}$ \citep{FeastWhitelock1997, Uemuraea2000, Eliasea2006} and "positive, negative and zero values" for the rotation gradient. Treating the Galactic potential locally as a Mestel potential, we obtain an epicycle frequency of 
\begin{align}
    \kappa = 41 \pm 2.5 \,\mathrm{km\,s}^{-1}{\mathrm{kpc}}^{-1}.
\end{align}For the AVR, \citet{AumerBinney2009} fitted $\sigma_R$ as a function of age $\tau$ with a power law:
\begin{align}
    \sigma_R(\tau) &= v_{10} \left(\frac{\tau + \tau_1}{10\,\mathrm{Gyr} + \tau_1}\right)^\beta, \\
    \beta &= 0.307, \\
    \tau_1 &= 0.001\,\mathrm{Gyr}, \\
    v_{10} &= 41.899\,\mathrm{km\,s}^{-1}.
\end{align}
They fitted the star formation rate history (SFR) as an exponential:
\begin{align}
    \mathrm{SFR}(\tau) &\propto \exp(\gamma \tau), \\
    \gamma &= 0.117 \mathrm{Gyr}^{-1}, \\
    \tau_\mathrm{max} &= 12.5 \mathrm{Gyr}.
\end{align}
This model is approximately the same as model $\mathrm{Y1}\xi-$ in \citet{Aumerea2016}, which provided a good fit to their measurements of $\sigma_R$ as a function of age. We use the total stellar surface density of the Solar Neighbourhood provided by \citet{BovyRix2013}:
\begin{align}
    \Sigma_\mathrm{stellar} = 38 \pm 4 \mathrm{M}_\odot\mathrm{\,pc}^{-2}.
\end{align}
For these properties of the Solar Neighbourhood we find
\begin{align}
    Q_\mathrm{\delta=5/3} &= 1.58, \\
    Q_\mathrm{Rk} &= 1.45.
\end{align}
Without full error propagation, we can demonstrate the uncertainty here by setting the gas surface density (which dominates the uncertainty) to its upper and lower bounds, giving respectively
\begin{align}
    Q_\mathrm{\delta=5/3} &= 1.06,\,2.13 \; , \\
    Q_\mathrm{Rk} &= 1.04,\,2.10 \; .
\end{align}
\citet{Aumerea2016} found in their N-body simulation with star formation that the disc settled at a $Q$ value of approximately 1.5, so both central values would be reasonable, while the upper and lower bounds of both appear to be unrealistic. To understand the relative contributions of stars and gas to the spiral structure, we write the dispersion relation (\ref{eq:DispersionRelationRk}) as a sum of polarisations as described in \citet{BinneyTremaine2008}:
\begin{align}
    \sum_{i=1}^{n_\mathrm{g}} \tilde{P}_{\mathrm{g}, i}(k, s) + \sum_{j=1}^{n_\mathrm{s}} \tilde{P}_{\mathrm{s}, j}(k, s) = 1,
\end{align}
\begin{align}
    \tilde{P}_{\mathrm{g}, i}(k, s) &= \frac{2 \pi G k \Sigma_{\mathrm{g}, i}}{\kappa^2(1 - s^2) + k^2 c_i^2}, \label{eq:gas_polarisation} \\
    \tilde{P}_{\mathrm{s}, j}(k, s) &= \frac{2 \pi G k \Sigma_{\mathrm{s}, j} \mathcal{F}(s, \chi_j)}{\kappa^2 (1 - s^2)}. \label{eq:stellar_polarisation}
\end{align}
These polarisations can be considered the relative contribution to a mode from each component. Figure \ref{fig:mw_dr_polarisations} shows the dispersion relation for the Solar Neighbourhood, with the total polarisations of stars and gas also plotted as a function of $k$. We see that stars tend to dominate the low-wavenumber modes while gas dominates the higher wavenumber-modes (the contributions will cross over again at extremely high $k$, but in this limit modes are simply coherent epicycles with no effect from self-gravity). Figure \ref{fig:m_amplification_errors} plots the maximum GLB swing amplification factor against the $m$ number of a spiral ($k_\mathrm{crit} R/X$) given this dispersion relation, with the total gas surface density taking its lower bound, central value and upper bound in the left, centre and right panels respectively (with all other parameters taking the central values of their uncertainty ranges). The total stellar and gas polarisations with $k = \sqrt{3/2} m/R$ are plotted against $m$: this is the value of $k$ at which the reduced spring rate $\tilde{\kappa}^2$ is minimised in \citet{Toomre1981}, so is likely to correspond to the pitch angle where amplification is strongest for a given $m$. Therefore the plotted polarisations are indicative of the stellar and gas contributions to swing amplification for a wave of the given $m$ number. We see that stars dominate swing amplification for $m \lesssim 10$, with gas dominant for $m \gtrsim 20$. Focusing on the middle panel, swing amplification under the GLB equation is strongest for $m$ between 10 and 25, but observations suggest an $m$ number of 4 for the milky way \citep{BobylevBajkova2014, ShenZheng2020}, with some authors suggesting $m = 2$ (e.g. \citealt{GrosbolCarraro2018}). This would suggest that, while higher $m$-numbers suffer the strongest swing amplification for a single wave, feedback mechanisms which converted amplified trailing waves into new leading waves might be more efficient for waves with a lower $m$-number (and therefore a greater radial range between their Lindblad resonances). The same conclusion can be drawn when we set the surface density of the gas to the lower or upper bound of its uncertainty range, as plotted in the left and right panels respectively. Our analysis would still favour $ m = 4$ waves over $m=2$ waves, as in all cases the maximum amplification factor at $m=2$ is less than 1.25. A feedback mechanism would have to be extremely efficient to convert a trailing wave which had been amplified by factor 1.25 into a leading wave of greater amplitude than the initial leading wave. With realistic feedback there can't be growth for $m=2$ waves over multiple swing amplification feedback cycles.

\begin{figure}
    \centering
    \includegraphics[width=\linewidth]{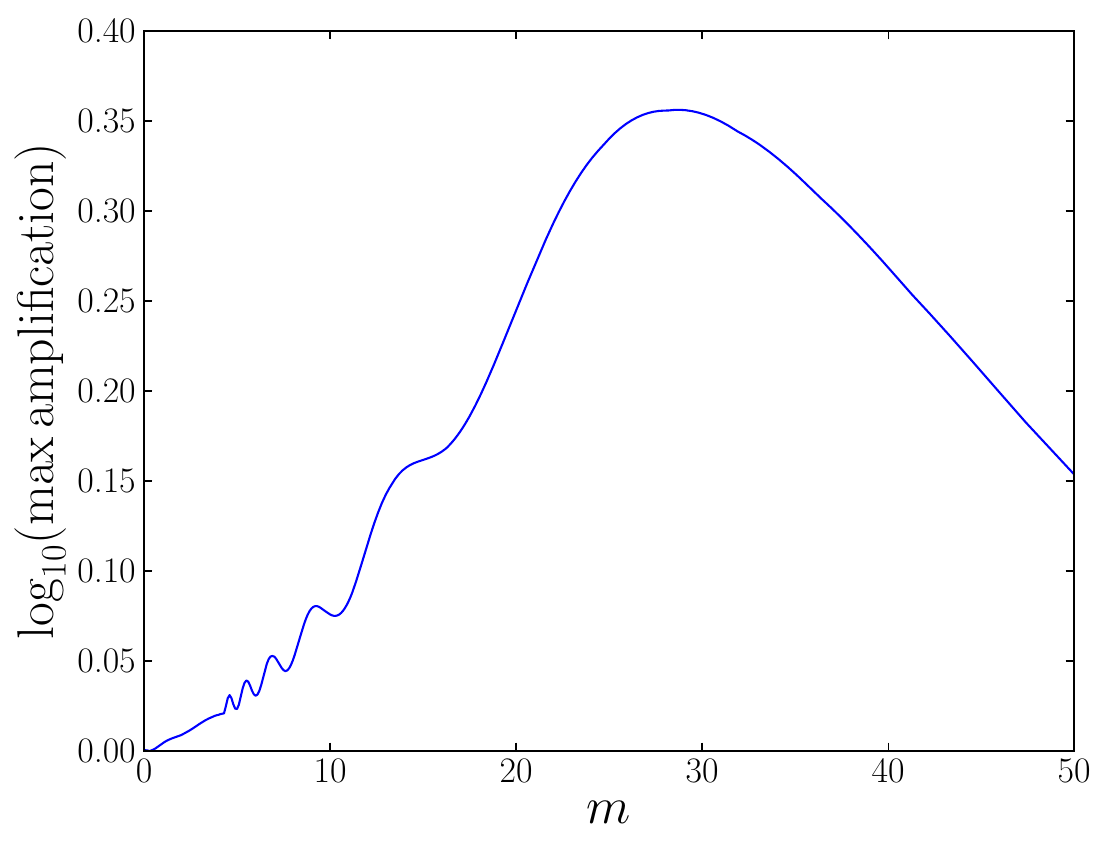}
    \caption{Maximum swing amplification factor obtainable under the GLB equation for a disc consisting only of the gas components of the Solar Neighbourhood with the properties listed in Table \ref{tab:gas_properties} and Section \ref{sec:solar_neighbourhood}}
    \label{fig:m_amplification_gas_only}
\end{figure}

We see in the middle panel of Figure \ref{fig:m_amplification_errors} that strong amplification with significant gas contribution can be obtained for $m$ value roughly between 10 and 25, without much variation in strength. The same can be said when the gas surface density is at its upper bound, shown in the right panel of Figure \ref{fig:m_amplification_errors}, although the gas surface density at its lower bound is insufficient (Left panel of Figure \ref{fig:m_amplification_errors}). Therefore (as long as the gas surface density isn't too low) we might expect to see significant flocculent spiral structure of this typical length scale in the ISM, from the swing amplification of strong signals such as the supernova remnants (structure which might be decomposed into leading and trailing spirals). Figure \ref{fig:m_amplification_gas_only} shows the maximum swing amplification factor (with gas surface density at its central value) as a function of $m$ for a disc consisting only of the gas components of the Solar Neighbourhood. The amplification factor is peaked at $m$ = 30, with relatively strong amplification for $m$ between 20 and 40. Spiral structure resulting from supernova remnants will be gas dominated, but may involve a response from the stellar disc, so both the middle panel of Figure \ref{fig:m_amplification_errors} and Figure \ref{fig:m_amplification_gas_only} are indicative of such structure.
\subsection{Application to M74} \label{sec:m74}

\begin{figure}
    \centering
    \includegraphics[width=\linewidth]{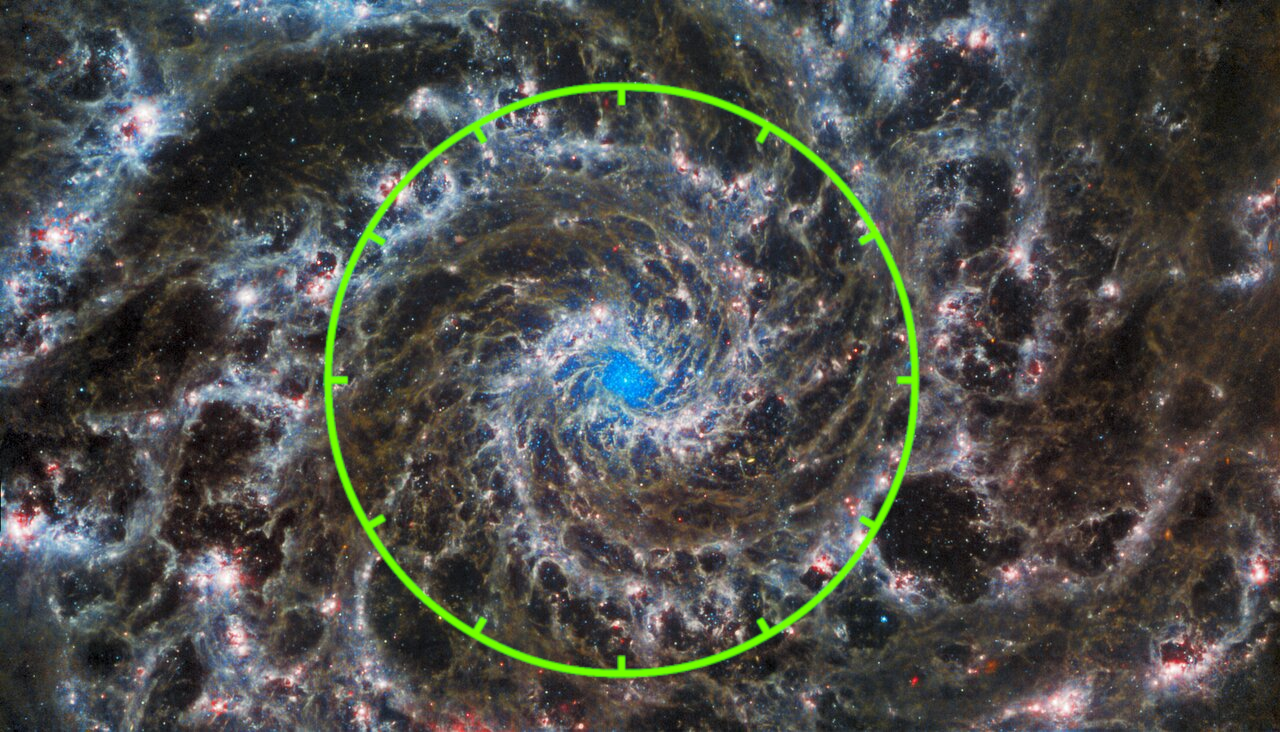}
    \caption{JWST/MIRI image of M74 [image credit: ESA/Webb, NASA \& CSA, J. Lee and the PHANGS-JWST Team. Acknowledgement: J. Schmidt]. We have overlaid a circle with 12 notches at a radius close to 2.6 kpc (the region for which we have listed properties in Table \ref{tab:gas_properties} and Section \ref{sec:m74}). Note the flocculent spiral structure in the ISM between the main spiral arms, which appears to have no dominant $m$-number but a range of wavelengths corresponding to $m$-numbers between about 10 and 25, as the GLB equation appears to predict (Figure \ref{fig:m_amplification_m74}, Figure \ref{fig:m_amplification_m74_gas_only}).}
    \label{fig:m74_jwst}
\end{figure}

\begin{figure}
    \centering
    \includegraphics[width=\linewidth]{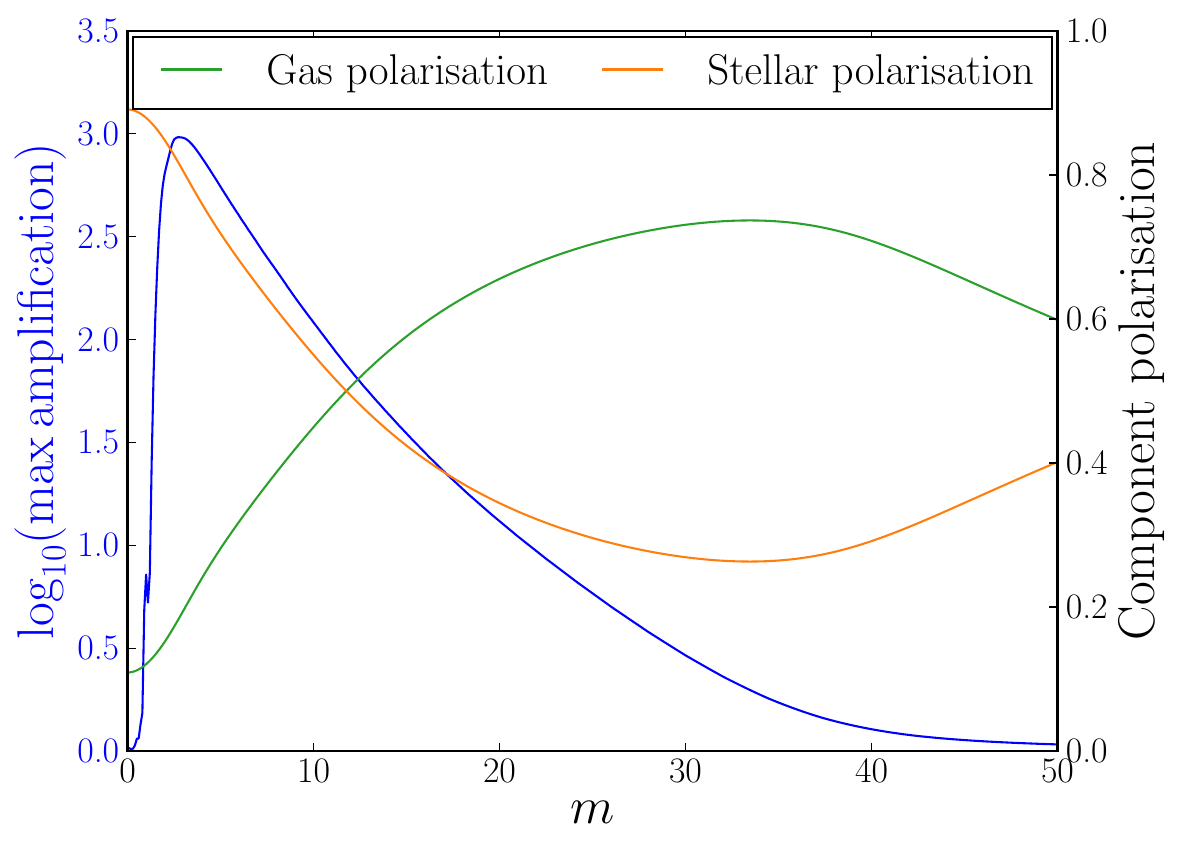}
    \caption{Maximum swing amplification factor obtainable under the GLB equation, as a function of $m$, given the properties of M74 at 2.6 kpc from the galactic centre listed in Section \ref{sec:m74} (include Table \ref{tab:m74_components}). Also plotted are the total polarisations of stellar and gas components, calculated at $k = \sqrt{3/2} k_\mathrm{crit}/X$ as in Figure \ref{fig:m_amplification_errors}. Note the strongest amplification is at $m \approx 3$, yet M74 has an $m=2$ spiral pattern, suggesting low $m$-numbers are preferred by the feedback mechanism of swing amplification. Note also that there is strong swing amplification and gas contribution for $m$ between 10 and 25, consistent with the flocculent spiral structure apparent in the ISM between the main spiral arms of M74 in Figure \ref{fig:m74_jwst}.}
    \label{fig:m_amplification_m74}
\end{figure}

\begin{figure}
    \centering
    \includegraphics[width=\linewidth]{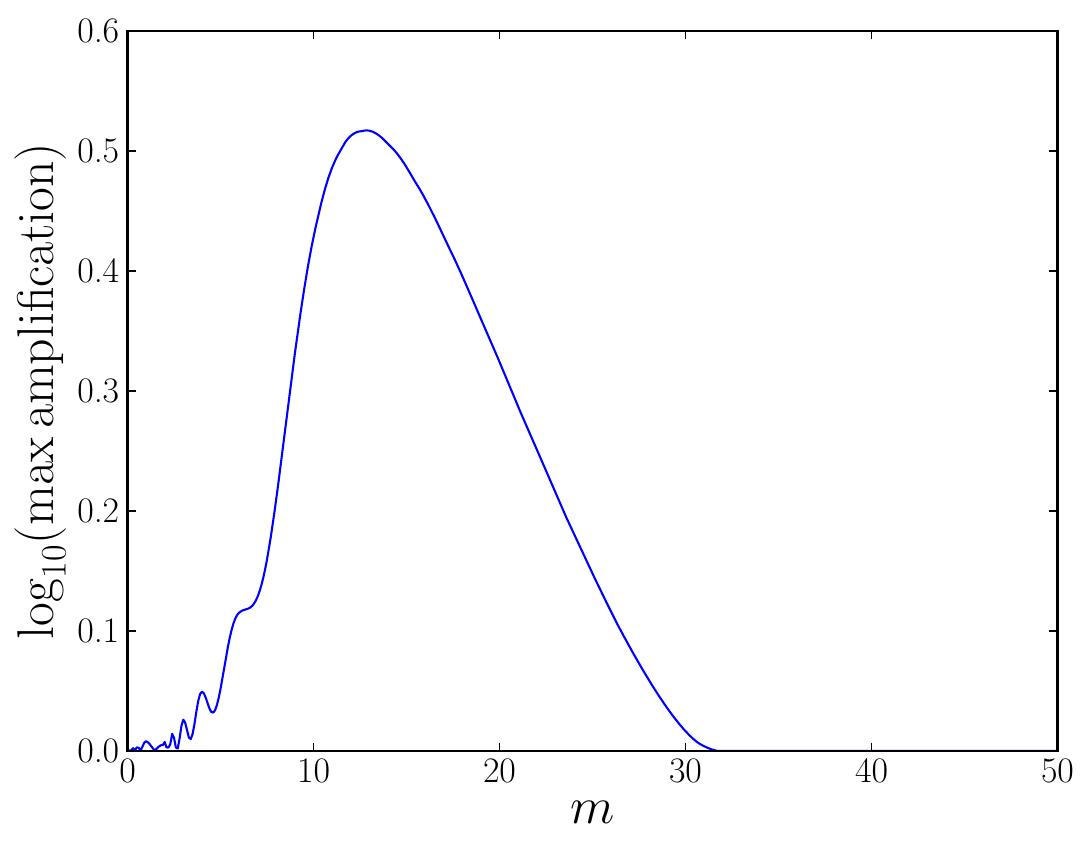}
    \caption{Maximum swing amplification factor obtainable under the GLB equation for a disc consisting only of the gas components of M74 (at 2.6 kpc from the galactic centre) with the properties listed in Table \ref{tab:m74_components} and Section \ref{sec:m74} Relatively strong amplification is predicted for $m$ between 10 and 25, suggesting flocculent spiral structure in the ISM as seen in Figure \ref{fig:m74_jwst}.}
    \label{fig:m_amplification_m74_gas_only}
\end{figure}

We can apply the same methods to external galaxies. M74 (NGC 0628), imaged by JWST/MIRI in Figure \ref{fig:m74_jwst}, is an $m = 2$ grand design spiral galaxy with apparent complicated spiral structure in the ISM between the main spiral arms. Table \ref{tab:m74_components} lists the surface densities and sound speeds/velocity dispersions of three components comprising the disc at 2.6 kpc from the galactic centre. \citet{Aniyanea2018} found a circular orbit speed of 150 $\mathrm{km\,s}^{-1}$ at 2.6 kpc from the galactic centre, corresponding to an epicycle frequency of $\kappa = 81.6 \mathrm{km\,s}^{-1} \mathrm{kpc}^{-1}$ if we assume a flat circular orbit speed curve. Because the galaxy is face-on, the rotation curve and so epicycle frequency have high uncertainty. Using this value for $\kappa$ and the component properties in Table \ref{tab:m74_components} we find that the disc is marginally unstable to axisymmetric perturbations, with a $Q_{\delta = 5/3}$ value of 0.99. Given the uncertainty on $\kappa$ is large, we choose a $\kappa$ value of 85 $\mathrm{km\,s}^{-1}\mathrm{kpc}^{-1}$, giving a $Q_{\delta=5/3}$ value of 1.03. Figure \ref{fig:m_amplification_m74} shows maximum amplification as a function of $m$, with polarisations plotted as in Figure \ref{fig:m_amplification_errors}. Amplification is maximal at $m = 3$ and stellar-dominated, but is still strong at $m = 2$, allowing an $m = 2$ spiral to dominate as seen in Figure \ref{fig:m74_jwst}. Strong amplification with significant gas contribution is predicted for $m$ between 10 and 25. Figure \ref{fig:m_amplification_m74_gas_only} shows maximum amplication as a function of $m$ for a disc consisting of only the gas component of M74 at 2.6kpc from the centre. Here amplification peaks at $m \approx 15$, with relatively strong amplification for $m$ between 10 and 25. Since spiral structure in the ISM would be given by the amplification of existing features (such as supernova remnants) rather than feedback cycle, the expected prevalence of spiral structure of a given $m$-number depends only on its amplification factor. Therefore the broad peak in Figure \ref{fig:m_amplification_m74_gas_only} implies that no particular $m$-number will be dominant. Of course other contributions will e.g. come from supernova feedback, which in turn, however, should again experience some swing amplification. Overall, our simplistic prediction provides an explanation for the apparent flocculent spiral structure in the ISM between the spiral arms in Figure $\ref{fig:m74_jwst}$, in which there seems to be no dominant $m$-number. 
\begin{table}
    \centering
    \begin{tabular}{|c|c|c|}
    \hline
        Description & $\Sigma_i/\mathrm{M}_\odot\mathrm{pc}^{-2}$ & $c_i/\mathrm{km\,s}^{-1}$ or $\sigma_{R, i}/\mathrm{km\,s}^{-1}$ \\
    \hline
        Gas & 35 & 9 \\
        Cold stars & 63 & 27.8 \\
        Hot stars & 223 & 92.3 \\
        \hline
    \end{tabular}
    \caption{Properties of the M74 components at 2.6 kpc from the galactic centre. All from \citet{Aniyanea2018}, except for the $c$ value for the gas, which is taken from \citet{ShostakVanderKruit1984} as a typical value near the galactic centre.}
    \label{tab:m74_components}
\end{table}
\section{Conclusions}
We have developed a more consistent framework to understand the disc instability criterion $Q$ for a disc of stars and gas, showing that pre-existing definitions are inconsistent. We proposed a re-definition of $Q$ that has close to a 1:1 relation with swing amplification, and applied it to the Milky Way and M74.

Fundamentally, $Q$  parametrises at least two different types of instability: $Q = 1$ is the boundary to axisymmetric instability \citep{Toomre1964}. However, Galactic discs commonly remain at $Q > 1$. In this regime, swing amplification (or related processes) drive spiral structure, holding discs in balance between star formation lowering $Q$ and heating from spiral patterns increasing $Q$ by raising the velocity dispersions of older populations. 
Regardless of being defined for the former, but not the latter, $Q$ has been made the standard currency for quantifying all spiral instabilities, even in multi-component discs. We have shown that this leads to inconsistent mapping between $Q$ and instability, particularly for discs with different gas fractions. 

We suggest a re-definition of $Q$ that preserves the axisymmetric instability limit at $Q = 1$, converges to the standard definition for a single-component (infinitely thin) isothermal disc, but meaningfully  quantifies spiral instability at $Q > 1$. To this end, we adopt the formal criterion of minimising the spread $\Delta_\mathrm{GLB}$ of maximum swing amplification (equation \ref{eq:delta_glb_definition} which uses the GLB equation (equation \ref{eq:glb_equation})) for discs of various compositions at each fixed $Q$. Most importantly, the pre-existing definition of gas component $Q$ needs to be approximately squared for consistency with the stellar $Q$. For multi-component discs we introduced $Q_{\delta=5/3}$ (equation \ref{eq:Q_delta_definition}), which significantly improves $\Delta_\mathrm{GLB}$ over any pre-existing definition. For vanishing gas fractions it reduces to $Q_\mathrm{Rk}$ \citep{Rafikov2001}, which we show to be the most consistent choice for stars-only discs (see Figure \ref{fig:avr_amp_dr}). We suggest $Q_{\delta=5/3}$ be used as the new standard definition for thin, multi-component discs, and provide code to calculate it. For convenience, where computing resources are limited, we have defined an approximation to $Q_\delta$, $Q_{\delta, \mathrm{RW}}$, as an iterative formula analogous to \citet{RomeoWiegert2011}.  $Q_{\delta, \mathrm{RW}}$ is typically acceptable for two-component discs but not for many components, as its error scales with $\sqrt{N_\mathrm{components}}$.

We cautiously apply the reduction factors from \citet{Romeo1992} (with concerns about the validity of their approximations) to the WKB dispersion relation to describe realistically thick two-component discs. Our new definition $Q_{\delta, \epsilon}$ (equation \ref{eq:q_delta_epsilon_definition}) with $\delta = 5/3$, $\epsilon \approx 2$ performs best on the amplification spread $\Delta_{\mathrm{GLB}}$, with the caveat that the validity of the approximation for finite thickness is doubtful. Nonetheless this $Q_{\delta, \epsilon}$ is the best available guess. It also correctly describes the axisymmetric stability criterion (given the WKB dispersion relation), and again reduces to the standard, thin, one-component $Q$. Unlike for $Q_{\delta}$ we found no simple approximation to the calculation-intensive $Q_{\delta, \epsilon}$.

There are central caveats to provide here: our case is firm for qualitatively changing the way we should define $Q$ and particularly for changing the relative definition between gas and stellar components. Quantitatively, our computations rely on an imperfect/incomplete corpus of mathematical formalism (GLB and JT approximations) with significant simplifying assumptions. However, at current stage, our $Q$ definition offers the most consistent map to instability and thus should be used until a better theoretical framework for spiral theory arises. In this case, our strategy can serve future papers.

We applied this to the Milky Way and M74. Our re-definition raises the $Q$ values contributed by the gas components. For the Solar Neighbourhood we obtain $Q_\mathrm{\delta=5/3} = 1.58$, which is consistent with simulations with continuous star formation \citep{Aumerea2016} and mildly larger than the previous best estimate $Q_\mathrm{Rk}=1.44$ \citep{Rafikov2001}. While the change is in the right direction, we note that for the Milky Way the error bars on $Q$ exceed this, due to the large uncertainty on the surface density of the ISM. On the other hand, the argument can be turned around: with improved understanding of the of behaviour of $Q$ in discs of stars and gas, one could in fact constrain ISM density and velocity dispersion/sound speed.

The maximum amplification factor as a function of $m$-number for the Solar Neighbourhood and M74 offers qualitative insights into their spiral structures. M74 has a lower $Q$ estimate (close to unity) than the Solar Neighbourhood and so a much higher maximum swing amplification value. M74's peak of amplification is at significantly lower $m$ compared to the Solar Neighbourhood. This coincides well with the current observational inference for the Milky Way: likely a not overly well-defined $m = 4$ (or similar) pattern, contrasting with M74's prominent two-arm ($m = 2$) grand design spiral (without any obvious influence by satellite galaxies). Curiously, the 
formalism predicts a peak of instability broad in $m$, while for both galaxies the stellar spiral patterns seem to favour the low-$m$ side. This re-emphasises the question how instability is seeded/propagated. Possibilities are that there are non-linear effects, or that specifics of feedback/groove instability might favour lower $m$-numbers. 

We also draw a tentative connection between the high $m$ substructure in M74 (Figure \ref{fig:m74_jwst}) and the local and other spurs \citep{Elmegreen1980, LaVigneea2006} in the Milky Way. The formalism here predicts that there is instability at higher $m$-numbers, which is increasingly dominated by the gas - above a certain $m$ stars only passively/hardly partake. Consistently in M74 JWST revealed a flocculent pattern in the gas/ISM between the main spiral arms with varying $m$-number of order $m \sim 20$, where the model predicts a wide peak of instability. The given parameters predict that the Milky Way should display a similar high-$m$ substructure, which will appear dominantly in the gas, but only weakly in the older stellar components.
\section*{Acknowledgements}
It is a pleasure to thank L. Athanassoula and D. Kawata for helpful and insightful discussions, and to thank W. Dehnen and R. Chiba for additionally providing comments on the paper draft. K.G. is funded by an STFC PhD studentship. R.S. gratefully acknowledges funding by a Royal Society University Research Fellowship.
\section*{Data Availability}
The data and code underlying this article will be shared on reasonable request to the corresponding author. A Jupyter notebook (Python) is provided for the calculation of our new definition $Q_\mathrm{\delta}$ for thin discs, as well as the existing definition $Q_\mathrm{Rk}$. Available at \url{https://github.com/kitgeorge/multi-component_Q}


\bibliographystyle{mnras}
\bibliography{references} 



\appendix
\section{\boldmath $\Delta_\mathrm{GLB}$ plots for our finite-thickness \boldmath $Q$ definition}
\begin{figure}
    \centering
    \includegraphics[width=0.5\textwidth]{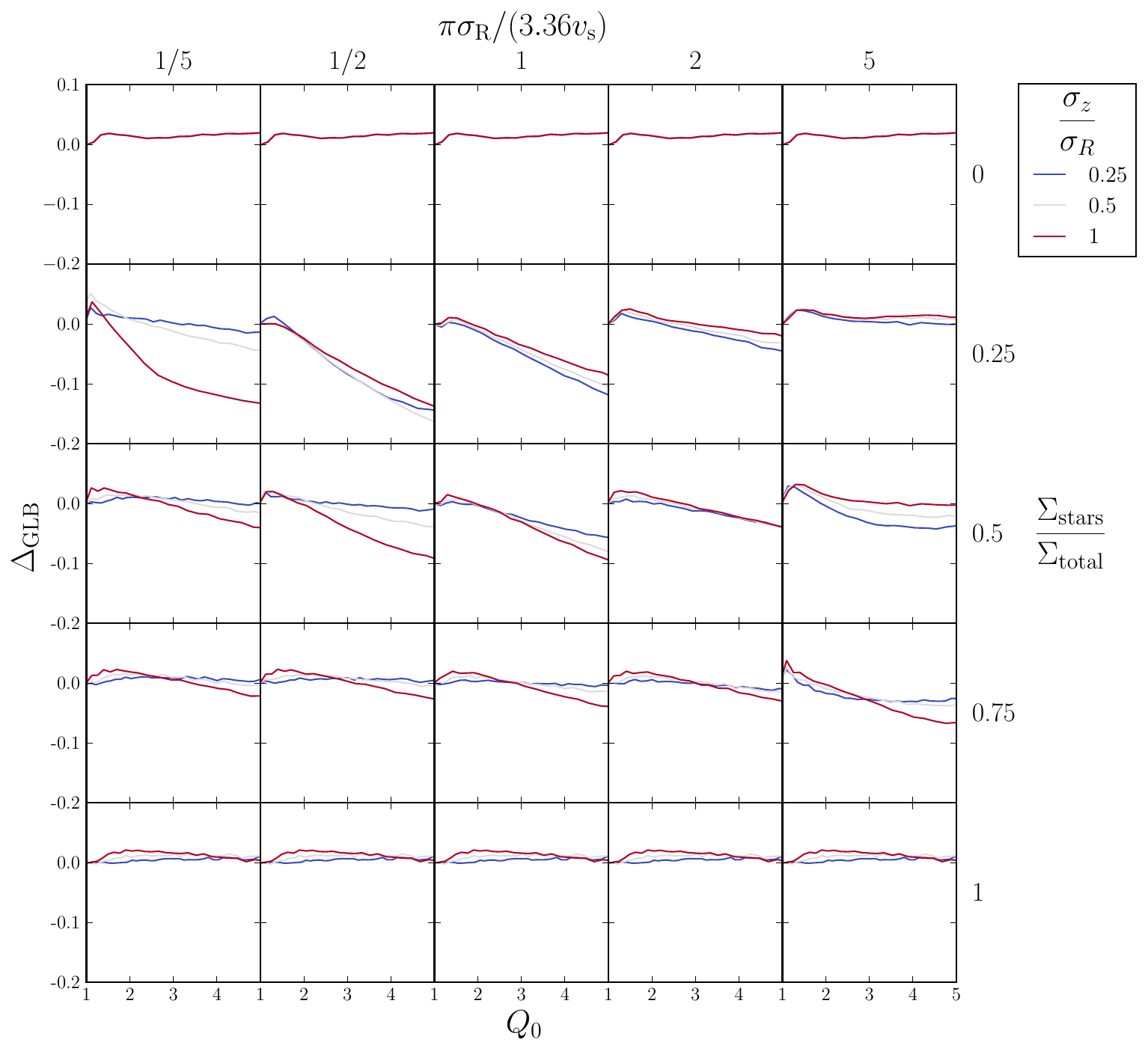}
    \caption{$\Delta_\mathrm{GLB}$ as a function of $Q_0$ for $Q_{\delta, \epsilon}$ with $\delta = \frac{5}{3}, \, \epsilon = \frac{11}{6}$ for two-component, realistically thick discs of stars and gas with various mass ratios and sound speed/velocity dispersion ratios between the two components and various vertical to radial velocity dispersion ratios for the stellar component. $Q_\mathrm{\delta, \epsilon}$ seems to be most consistent in terms of $\Delta_\mathrm{GLB}$ when $\delta = 5/3$, $\epsilon$ between 11/6 and 2 (Figure \ref{fig:epsilon=2}).}
    \label{fig:epsilon=11/6}
\end{figure}
\begin{figure}
    \centering
    \includegraphics[width=0.5\textwidth]{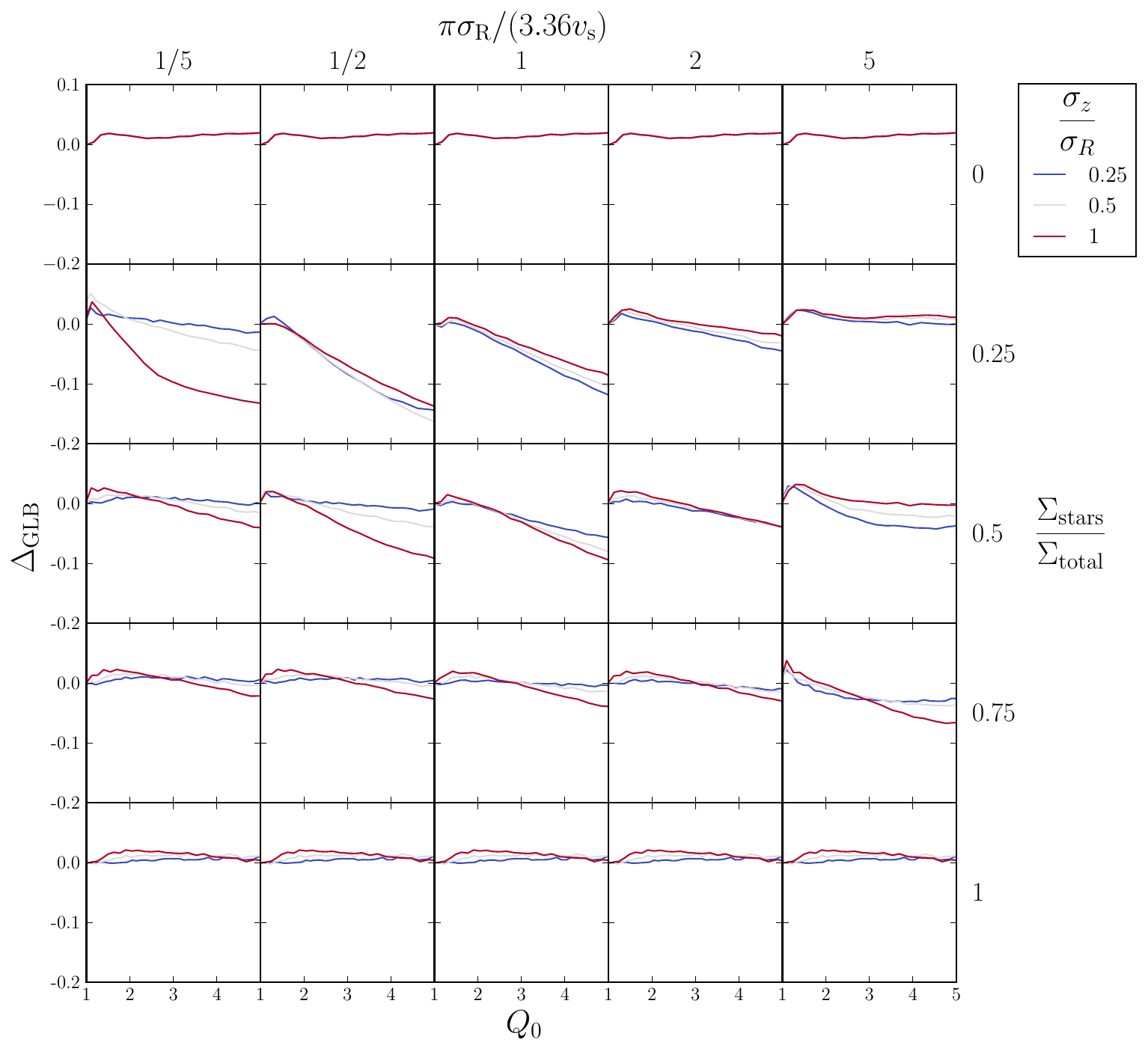}
    \caption{$\Delta_\mathrm{GLB}$ as a function of $Q_0$ for $Q_{\delta, \epsilon}$ with $\delta = \frac{5}{3}, \epsilon = 2$ for two-component, realistically thick discs of stars and gas with various mass ratios and sound speed/velocity dispersion ratios between the two components and various vertical to radial velocity dispersion ratios for the stellar component. $Q_\mathrm{\delta, \epsilon}$ seems to be most consistent in terms of $\Delta_\mathrm{GLB}$ when $\delta = 5/3$, $\epsilon$ between 11/6 (Figure \ref{fig:epsilon=11/6}) and 2 .}
    \label{fig:epsilon=2}
\end{figure}

\bsp	
\label{lastpage}
\end{document}